\newif\ifemulate
\documentclass{emulateapj}  \usepackage{apjfonts} \emulatetrue

\slugcomment{To appear in ApJ September 20, 2008}

\newlength{\figsize}
\ifemulate
\else 
\fi
\newcommand{\Figure}[2]
           {\centering\leavevmode\includegraphics[width=#2,clip]{#1.ps}}
\newcommand{\eq}[1]{\begin{equation} #1 \end{equation}}

\newcommand\about  {\hbox{$\sim$}}
\renewcommand\deg  {\hbox{$^\circ$}}
\newcommand{\E}[1]{\hbox{$10^{#1}$}}

\newcommand\x      {\hbox{$\times$}}
\newcommand\half   {\hbox{$\frac12$}}
\newcommand\mic    {\hbox{$\mu$m}}

\newcommand\LEdd   {\hbox{$L_{\rm Edd}$}}
\newcommand\Mo     {\hbox{$M_{\odot}$}}

\newcommand\mH     {\hbox{$m_{\rm H}$}}

\newcommand\Nc     {\hbox{$N_{\rm C}$}}
\newcommand\N      {\hbox{$\cal N$}}
\newcommand\NT     {\hbox{${\cal N}_{\rm T}$}}
\newcommand\No     {\hbox{${\cal N}_0$}}
\newcommand\NH     {\hbox{$N_{\rm H}^{(1)}$}}
\newcommand\Ntor   {\hbox{$N_{\rm torus}^{\rm (eq)}$}}
\newcommand\Mtor   {\hbox{$M_{\rm torus}$}}
\newcommand\ntot   {\hbox{$n_{\rm tot}$}}
\newcommand\Pesc   {\hbox{$P_{\rm esc}$}}
\newcommand\Rd     {\hbox{$R_{\rm d}$}}
\newcommand\Ro     {\hbox{$R_{\rm o}$}}
\newcommand\Rc     {\hbox{$R_{\rm c}$}}
\newcommand\Rx     {\hbox{$R_{\rm x}$}}
\newcommand\Sl     {\hbox{$S_{\rm c,\lambda}$}}
\newcommand\tl     {\hbox{$\tau_{\rm \lambda}$}}
\newcommand\tV     {\hbox{$\tau_V$}}
\newcommand\td     {\hbox{$\theta_{\rm d}$}}
\newcommand\cs     {\hbox{cm$^{-2}$}}
\newcommand\Ftor   {\hbox{$F_{\rm tor}$}}
\newcommand\FAGN  {\hbox{$F_{\rm AGN}$}}
\newcommand\pAGN  {\hbox{$p_{\rm AGN}$}}
\newcommand\IC    {\hbox{$I^{\rm C}_{\lambda}$}}
\newcommand\FC    {\hbox{$F^{\rm C}_{\lambda}$}}
\newcommand\Fcont {\hbox{$F_{\rm c,\lambda}$}}
\newcommand\erg   {\hbox{erg\,s$^{-1}$}}
\newcommand\kms   {\hbox{km\,s$^{-1}$}}
\newcommand\MBH   {\hbox{$M_{\bullet\,7}$}}
\newcommand\rpc   {\hbox{$r_{\rm pc}$}}
\newcommand\Ssil  {\hbox{$S_{10}$}}

 \shorttitle{AGN Dusty Tori: II. Observational Implications}
 \shortauthors{Nenkova et al}

\begin{document}

\title{AGN Dusty Tori: II. Observational Implications of Clumpiness}

\author{Maia Nenkova\altaffilmark{1},
        Matthew M. Sirocky \altaffilmark{2},
        Robert Nikutta\altaffilmark{2},
        \v{Z}eljko Ivezi\'c\altaffilmark{3}
        and Moshe Elitzur\altaffilmark{2}}

% \email{maia.nenkova@senecac.on.ca;
%        ivezic@astro.washington.edu;
%        sirockmm@pa.uky.edu;
%        robert@pa.uky.edu;
%        moshe@pa.uky.edu}

\altaffiltext{1}{Seneca College, 1750 Finch Ave. East, Toronto,
                 ON M2J 2X5, Canada; maia.nenkova@senecac.on.ca}
\altaffiltext{2}{Department of Physics and Astronomy, University of Kentucky,
                Lexington, KY 40506-0055; sirockmm@pa.uky.edu,
                robert@pa.uky.edu, moshe@pa.uky.edu}
\altaffiltext{3}{Department of Astronomy, University of Washington,
                 Seattle, WA 98105; ivezic@astro.washington.edu}

\begin{abstract}

From extensive radiative transfer calculations we find that clumpy torus models
with \No\ \about\ 5--15 dusty clouds along radial equatorial rays successfully
explain AGN infrared observations. The dust has standard Galactic composition,
with individual cloud optical depth \tV\,\about\ 30--100 at visual. The models
naturally explain the observed behavior of the 10\mic\ silicate feature, in
particular the lack of deep absorption features in AGN of any type. The weak
10\mic\ emission feature tentatively detected in type 2 QSO can be reproduced
if in these sources \No\ drops to \about 2 or \tV\ exceeds \about 100. The
clouds angular distribution must have a soft-edge, e.g., Gaussian profile, the
radial distribution should decrease as $1/r$ or $1/r^2$. Compact tori can
explain all observations, in agreement with the recent interferometric evidence
that the ratio of the torus outer to inner radius is perhaps as small as \about
5--10. Clumpy torus models can produce nearly isotropic IR emission together
with highly anisotropic obscuration, as required by observations. In contrast
with strict variants of unification schemes where the viewing-angle uniquely
determines the classification of an AGN into type 1 or 2, clumpiness implies
that it is only a probabilistic effect; a source can display type 1 properties
even from directions close to the equatorial plane. The fraction of obscured
sources depends not only on the torus angular thickness but also on the cloud
number \No. The observed decrease of this fraction at increasing luminosity can
be explained with a decrease of either torus angular thickness or cloud number,
but only the latter option explains also the possible emergence of a 10\mic\
emission feature in QSO2. X-ray obscuration, too, has a probabilistic nature.
Resulting from both dusty and dust-free clouds, X-ray attenuation might be
dominated by the dust-free clouds, giving rise to the observed type 1 QSO that
are X-ray obscured. Observations indicate that the obscuring torus and the
broad line region form a seamless distribution of clouds, with the transition
between the two regimes caused by dust sublimation. Torus clouds may have been
detected in the outflow component of H$_2$O maser emission from two AGN. Proper
motion measurements of the outflow masers, especially in Circinus, are a
promising method for probing the morphology and kinematics of torus clouds.

\end{abstract}


\section{INTRODUCTION}

Recent VLTI interferometric observations in the 8--13~\mic\ wavelength range by
\cite{Tristram07} confirm the presence of a geometrically thick, torus-like
dust distribution in the nucleus of Circinus, as required by unification
schemes of Seyfert galaxies. Several aspects of their data require that this
torus is irregular, or clumpy, in agreement with the earlier prediction of
\cite{Krolik88}.

We have recently developed the first formalism for handling clumpy AGN tori and
presented initial results \citep{NIE02, Elitzur04, Elitzur06, Elitzur07}. The
reported clumpy models have since been employed in a number of observational
studies, including the first analysis of Spitzer observations by the GOODS
Legacy project \citep{Treister04}. Our clumpy torus models were also employed
in the analysis of spatially-resolved, near-diffraction-limited 10~\mic\
spectra of the NGC~1068 nucleus \citep{Mason06}. The geometry and kinematics of
both water maser \citep{Greenhill97, Gallimore01} and narrow-line emission
\citep{Crenshaw00} indicate that the NGC~1068 torus and accretion disk are
oriented nearly edge-on. The \citeauthor{Mason06} clumpy model for IR emission
is the first to correctly reproduce the observed near-IR flux with an edge-on
orientation. In contrast, smooth-density models require viewing angles
22\deg--30\deg\ above the equatorial plane in order to bring into view the warm
face of the torus backside \citep{Granato97, Gratadour03, Fritz06}. Clumpiness
is also essential for understanding the puzzling interferometry result that
dust temperatures as different as $\ga$ 800 K and \about 200--300\,K are found
at such close proximity to each other \citep{Schartmann05}. The mounting
observational evidence in favor of clumpy, rather than smooth, dust
distribution in AGN tori has sparked additional modeling efforts by
\cite{Dullemond05} and \cite{Hoenig06}.

This two-paper series expands the analysis of \cite{NIE02}. In its first part
\cite[part I hereafter]{AGN1} we develop the full formalism for continuum
emission from clumpy media and construct the source functions of dusty
clouds---the building blocks of the AGN torus. Here we assemble these clouds
into complete models of the torus, and study the model predictions and their
implications to IR observations. In comparing the predictions of any torus
model with observations one faces a difficult problem---the overwhelming
majority of these observations do not properly isolate the torus IR emission.
Starburst emission is increasingly recognized as an important component of the
IR flux measured in many, perhaps most, AGN \citep[e.g.,][]{Netzer07}.  In
addition to this well known contamination, even IR from the immediate vicinity
of the AGN may not always originate exclusively from the torus, further
complicating modeling efforts. A case in point is the \cite{Mason06} modeling
of NGC1068. All flux measurements with apertures $< 0.5''$ are in good
agreement with the model results, but the flux collected with larger apertures
greatly exceeds the model predictions at wavelengths longer than \about 4\mic.
This discrepancy can be attributed to IR emission from nearby dust outside the
torus. Mason et al show that the torus contributes less than 30\% of the 10
\mic\ flux collected with apertures $\ge 1''$ and that the bulk of the
large-aperture flux comes at these wavelengths from dust in the ionization
cones; while less bright than the torus dust, it occupies a much larger volume
\citep[see also][]{Poncelet07}. On the other hand, the torus dominates the
emission at short wavelengths; at 2 \mic, more than 80\% of the flux measured
with apertures $\ge 1''$ comes from the torus even though its image size is
less than 0.04$''$ \citep{Weigelt04}.

These difficulties highlight a problem that afflicts all IR studies of AGN. The
torus emission can be expected to dominate the AGN observed flux at near IR
because such emission requires hot dust that exists only close to the center.
But longer wavelengths originate from cooler dust, and the torus contribution
can be overwhelmed by the surrounding regions. Unfortunately, there are not too
many sources like NGC1068. No other AGN has been observed as extensively and
almost no other observations have the angular resolution necessary to identify
the torus component, making it impossible to determine in any given source
which are the wavelengths dominated by torus emission. There are no easy
solutions to this problem. One possible workaround is to forgo fitting of the
spectral energy distribution (SED) in individual sources and examine instead
the observations of many sources to identify characteristics that can be
attributed to the torus signature. One example for the removal of the starburst
component is the \cite{Netzer07} composite SED analysis of the \emph{Spitzer}
observations of PG quasars. Netzer et al identify two sub-groups of ``weak
FIR'' and ``strong FIR'' QSOs and a third group of far-IR (FIR) non-detections.
Assuming a starburst origin for the far-IR, they subtract a starburst template
from the mean SED of each group. The residual SEDs are remarkably similar for
all three groups, and thus can be reasonably attributed to the intrinsic AGN
contribution, in spite of the many uncertainties. However, while presumably
intrinsic to the AGN, it is not clear what fraction of this emission originates
from the torus as opposed to the ionization cones. An example of sample
analysis that may have identified the torus component is the \cite{Hao07}
compilation of \emph{Spitzer} IR observations. In spite of the large aperture
of these measurements, Seyfert 1 and 2 galaxies show a markedly different
behavior for the 10\mic\ feature, both in their mean IR SEDs and in their
distributions of feature strength. Furthermore, Ultraluminous IR Galaxies
(ULIRG) that are not associated with AGN show yet another, entirely different
behavior, indicating that the observed mean behavior of Seyfert galaxies is
intrinsic to the AGN. Accepting the framework of the unification scheme, the
differences Hao et al find between the appearances of Seyfert 1 and 2 can be
reasonably attributed to the torus contribution; the ionization cones dust is
optically thin, therefore its IR emission is isotropic and cannot generate the
observed differences between types 1 and 2.

Here we invoke both approaches in comparing our model predictions with
observations. We start by assembling dusty clouds into complete models of the
torus, as described in \S\ref{sec:Model}. Our model predictions for torus
emission and the implications to IR observations are presented in
\S\S\ref{sec:SEDs}--\ref{sec:SPP}, while in \S\ref{sec:Others} we discuss
aspects of clumpiness that are unrelated to the IR emission, such as the torus
mass, unification statistics, etc. In \S\ref{sec:Discussion} we conclude with a
summary and discussion.

\section{MODEL OF A CLUMPY TORUS}
\label{sec:Model}

Consider an AGN with bolometric luminosity $L$ surrounded by a toroidal
distribution of dusty clouds (Fig.\ \ref{Fig:torus}). The ``naked'' AGN flux at
distance $D$ is $\FAGN = L/4\pi D^2$ at any direction, but because of
absorption and re-emission by the torus clouds the actual flux distribution is
anisotropic, with the level of anisotropy strongly dependent on wavelength.
The grain mix has standard interstellar properties (see paper I,
\S3.1.1 for details), the optical depth of each cloud is \tV\ at visual.

%\newpage

\subsection{Dust Sublimation}
\label{sec:sublimation}

The distribution inner radius \Rd\ is set by dust sublimation at temperature
$T_{\rm sub}$. From \S3.1.2 in part I,
\eq{\label{eq:Rd}
    \Rd \simeq 0.4\left(L\over\E{45}\,{\rm erg\,s^{-1}}\right)^{\!\!1/2}
                  \left(1500\,{\rm K} \over T_{\rm sub}\right)^{\!\!2.6}
                  \ \rm pc.
}
\cite{Barvainis87} derived an almost identical relation for \Rd. His eq.\ 5 has
the same normalization and only a slight difference in the power of $T_{\rm
sub}$ (2.8 instead of 2.6); this difference reflects the more detailed
radiative transfer calculations we perform. Here the distance \Rd\ is
determined from the temperature on the illuminated face of an optically thick
cloud of composite dust representing the grain mixture. The sharp boundary we
employ is an approximation. In reality, the transition between the dusty and
dust-free environments is gradual because individual components of the mix
sublimate at slightly different radii, with the largest grains surviving
closest to the AGN \citep{Schartmann05}. From near-IR reverberation
measurements, \cite{Minezaki04}
and \cite{Suganuma06} find that the inner radius of the dusty region is indeed
proportional to $L^{1/2}$, but the time lags they report are \about 2--3 times
shorter than predicted by eq.\ \ref{eq:Rd}. While this equation gives the
smallest radius at which the dust absorption coefficient reflects the full
grain mixture, the largest grains survive to closer radii, where they are
presumably detected by the reverberation measurements.

\ifemulate
%%%%%%%%%%%%%%%%%%%% Torus - 2 types %%%%%%%%%%%%%%%%%%%%%%%%%%%%%%%%
\begin{figure}
 \Figure{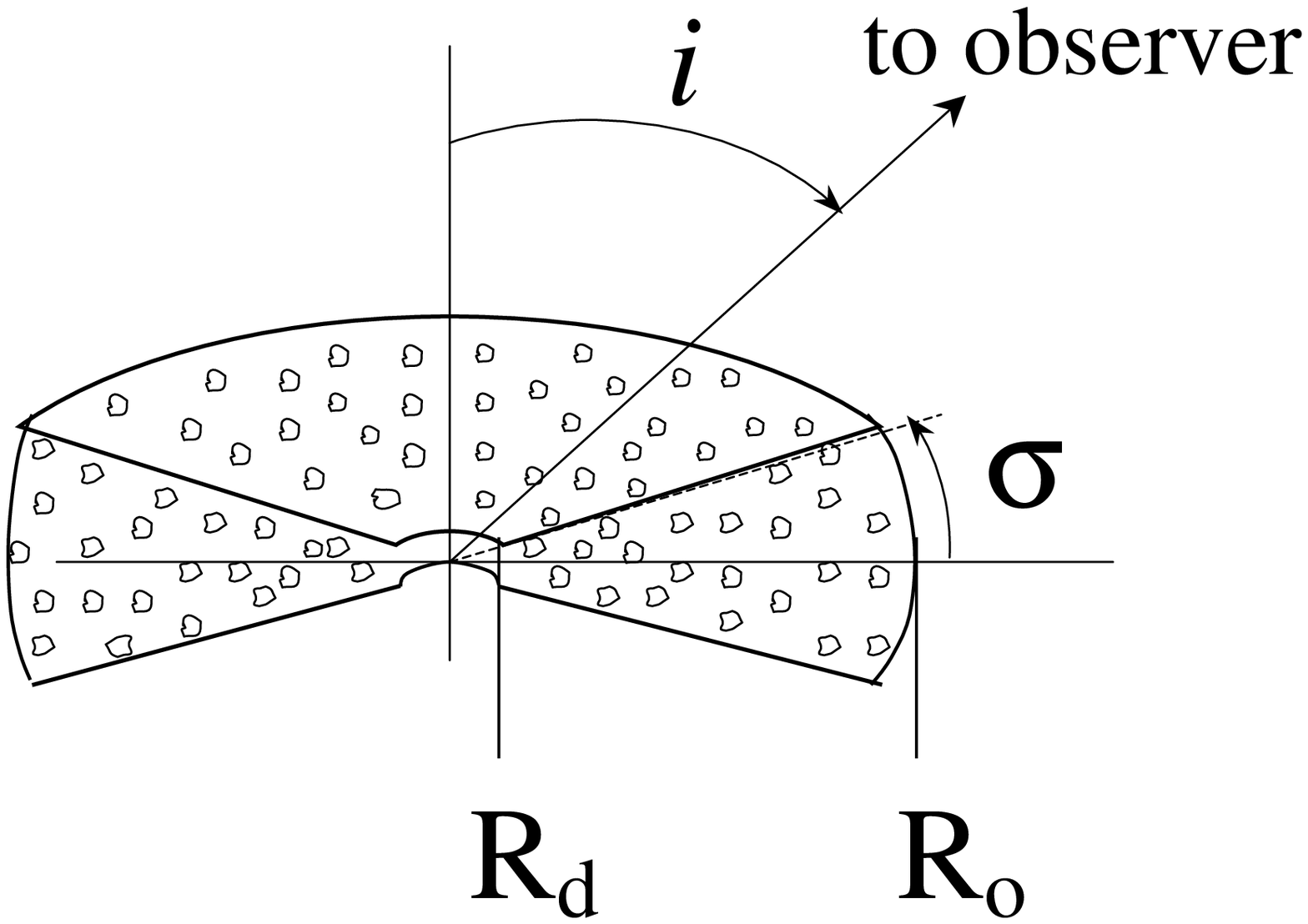}{0.49\figsize} \hfill
 \Figure{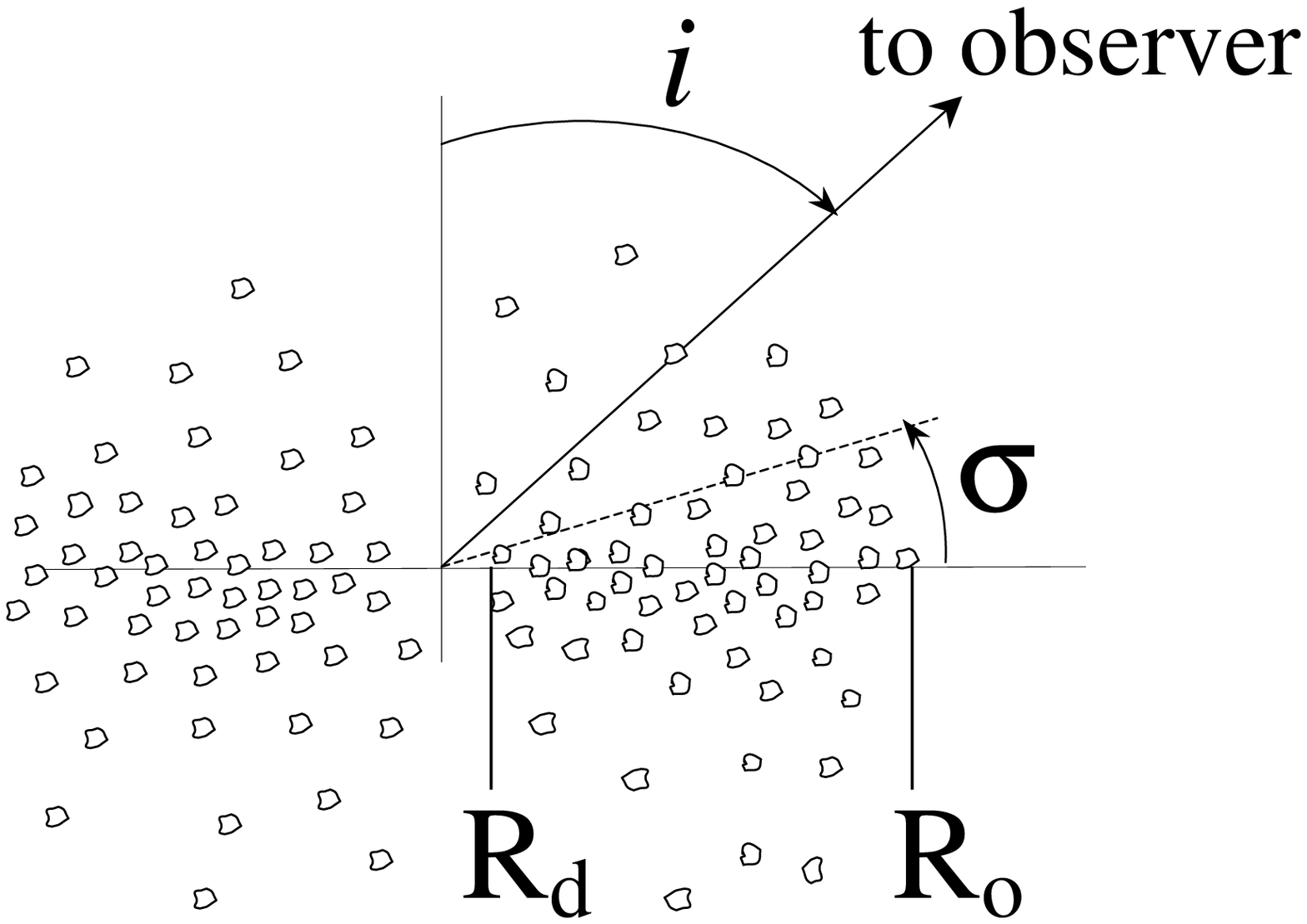}{0.49\figsize}

\caption{Model geometry: Dusty clouds, each with an optical depth \tV\ at
visual, occupy a toroidal volume from inner radius \Rd, determined by dust
sublimation (eq.\ \ref{eq:Rd}), to outer radius $\Ro = Y\Rd$. The radial
distribution is a power law $r^{-q}$, the total number of clouds along a radial
equatorial ray is \No. Various angular distributions, characterized by a width
parameter $\sigma$, were considered. The angular distribution has a sharp-edge
on the left and a smooth boundary (e.g., a Gaussian) on the right.}
\label{Fig:torus}
\end{figure}
%%%%%%%%%%%%%%%%%%%%%%%%%%%%%%%%%%%%%%%%%%%%%%%%%%%%%%%%%%%%%%
\fi

\subsection{The Cloud Distribution}
The torus extends radially out to $\Ro = Y\Rd$, with $Y$ a free parameter. The
total number of clouds, on average, along any radial equatorial ray is
specified by the parameter \No. We studied various forms for the variation of
$\NT(\beta)$, the total number of clouds along rays at angle $\beta$ from the
equator. Figure \ref{Fig:torus} shows on the left a sharp-edge uniform
distribution with $\NT(\beta) = \No$ within the angular width $|\beta| \le
\sigma$. In a Gaussian distribution, $\NT(\beta) = \
\No\exp(-\beta^2/\sigma^2)$.

The emission from the clumpy torus is found by integration along paths through
the cloud distribution (eq.\ 5, part I). Some of the computation technicalities
are described in the Appendix. The calculation requires the single cloud source
function \Sl, derived in part I, and the number of clouds per unit length,
$\Nc(r,\beta)$, as a function of $\beta$ and radial distance $r$. For this
distribution we assume a separable function with power law radial behavior
$r^{-q}$ so that
\eq{\label{eq:Nc}
    \Nc(r,\beta) = C\,{\NT(\beta)\over\Rd}\x\left({\Rd\over r}\right)^{\!\! q}
}
where $C = (\int_1^Y dy/y^q)^{-1}$ is a dimensionless constant (for a given
$Y$ and $\sigma$), ensuring the normalization $\NT(\beta) = \int\Nc(r,\beta) dr$.

\ifemulate
%%%%%%%%%%%%%%% Pesc  %%%%%%%%%%%%%%
\begin{figure}
 \Figure{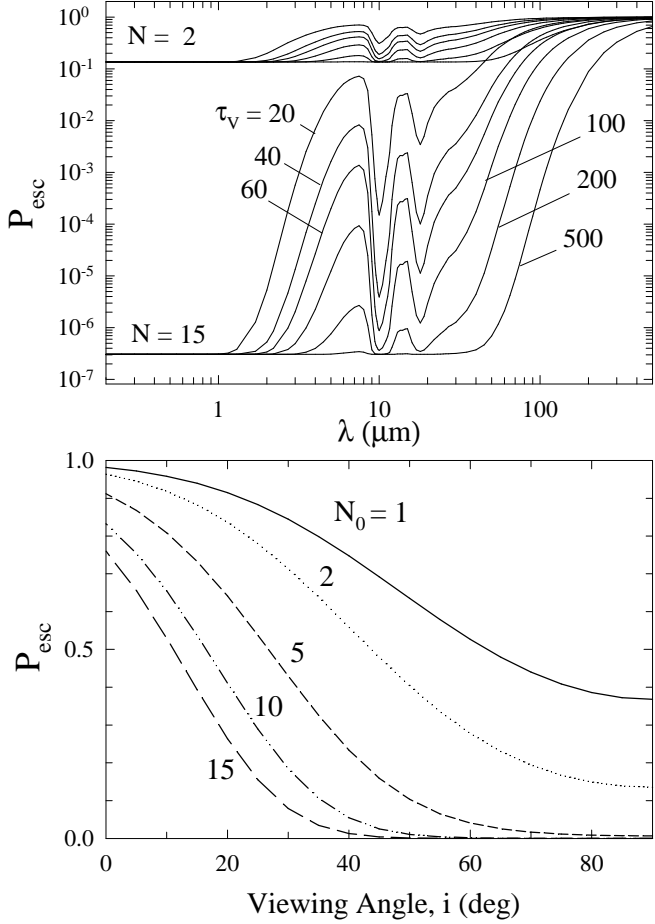}{ \figsize}

\caption{Behavior of the probability for photon escape along a path containing
\N\ clouds (eq.\ 4 part I). {\em Upper panel}: Wavelength variation of \Pesc\
for the indicated \N\ when the single cloud optical depth is \tV\ at visual.
{\em Lower panel}: The probability for an AGN photon to escape through the
torus in direction $i$ from the pole when each cloud is optically thick; this
is also the probability for unobscured view of the AGN at viewing angle $i$.
The total number of clouds varies according to $\NT(\beta) = \
\No\exp(-\beta^2/\sigma^2)$, where $\beta = \frac12\pi - i$ is angle from the
equatorial plane, with  $\sigma = 45\deg$ and \No\ as marked.}
 \label{Fig:Pesc}
\end{figure}

%%%%%%%%%%%%%%%%%%%%%%%%%%%%%%%%%%%%%%%%%%%%%%%%%%%%%%%
\fi

The observed torus radiation is affected not only by the emission from
individual clouds but also by the probability that emitted photons escape
through the rest of the path. The escape probability, \Pesc, is given in eq.\ 4
part I. For an overall number of clouds \N\ along a path, $\Pesc \simeq
\exp(-\N\tl)$ at wavelengths in which the optical depth of a single cloud obeys
$\tl < 1$, and $\Pesc \simeq \exp(-\N)$ when $\tl > 1$. Many of the detailed
results presented below can be readily understood from the dependence of \Pesc\
on wavelength and on torus viewing angle, shown in figure \ref{Fig:Pesc}.

\subsection{Scaling}
\label{sec:Scaling}

Because of general scaling properties of radiatively heated dust \citep{IE97},
the only effect of the overall luminosity is in setting up the bolometric flux
\FAGN\ and the dust sublimation radius \Rd\ (eq.\ \ref{eq:Rd}). For a given
torus model, the distributions of dust temperature and of brightness are unique
functions of the scaled radial distance $r/\Rd$: two sources with the same
cloud properties but different luminosities will have the same distributions in
terms of $r/\Rd$, only the more luminous one will have its brightness spread
over a larger area because of its larger \Rd\ (this point is explained further
in the Appendix). Denoting the torus flux by $F_{\lambda}$, the spectral shape
$F_{\lambda}/\FAGN$ is independent of $L$. The dependence of the torus SED on
the spectral shape of the AGN input radiation is limited to scattering
wavelengths, disappearing altogether at $\lambda \ga$ 2--3 \mic. There is a
similarly weak dependence on $T_{\rm sub}$. The output spectrum depends
primarily on \tV\ and the cloud distribution. Although the luminosity does not
affect the radiative transfer, it is entirely possible for torus properties to
be correlated with $L$ for some other reasons (e.g. $\sigma$, as in the
receding torus model).

%\newpage
\subsection {The AGN Contribution}
\label{sec:the_AGN}

In most figures we show only the contribution of the torus emission. However,
since the medium is clumpy, there is always a finite probability for an
unobscured view of the AGN, irrespective of the viewing angle. Because of the
probabilistic nature of the problem it is only possible to display the emerging
spectral shape with or without the AGN contribution and the probability for
each case (see bottom panel of figure \ref{Fig:Pesc}).

\ifemulate

%%%%%%%%%%%%%%% Models of sharp edge torus vs. Gaussian dsitribution %%%%%%%%%%
\begin{figure}
 \Figure{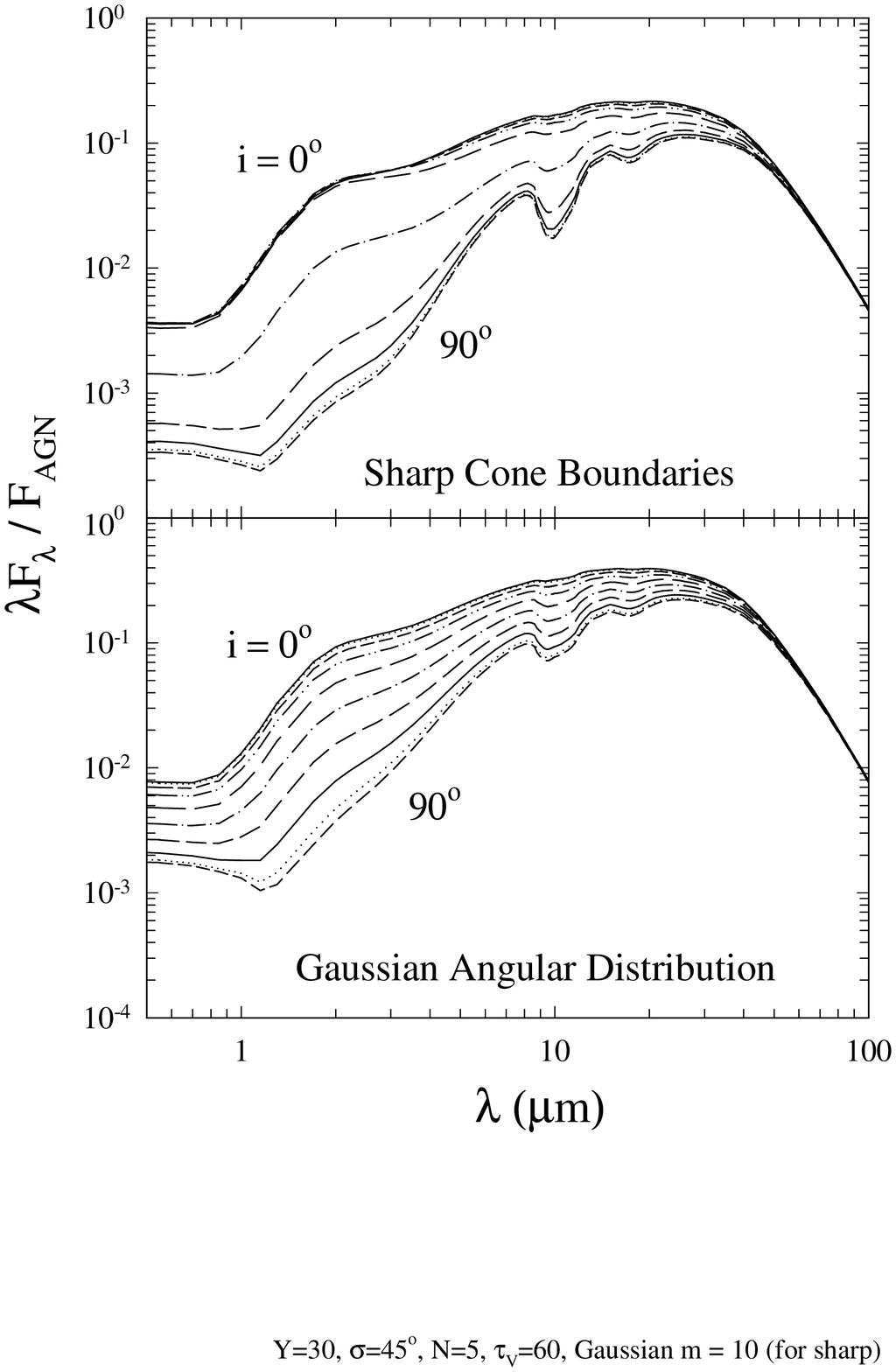}{\figsize}
\caption{Model spectra for a torus of clouds, each with optical depth \tV\ =
60. Radial distribution with $q = 1$ out to $Y$ = 30, with \No\ = 5 clouds
along radial equatorial rays (see eq.\ \ref{eq:Nc}). The angular distribution is
sharp-edged in the top panel, Gaussian in the bottom one (cf.\ fig.
\ref{Fig:torus}); both have a width parameter $\sigma$ = 45\deg. Different
curves show viewing angles that vary in 10\deg\ steps from pole-on ($i$ =
0\deg) to edge-on ($i$ = 90\deg). Fluxes scaled with $\FAGN = L/4\pi D^2$.}
\label{Fig:srpGauss}
\end{figure}
%%%%%%%%%%%%%%%%%%%%%%%%%%%%%%%%%%%%%%%%%%%%%%%%%%%%%%%

%%%%%%%%%%%%%%% Models vs. Data for Type 1,2 %%%%%%%%%%
\begin{figure}
 \Figure{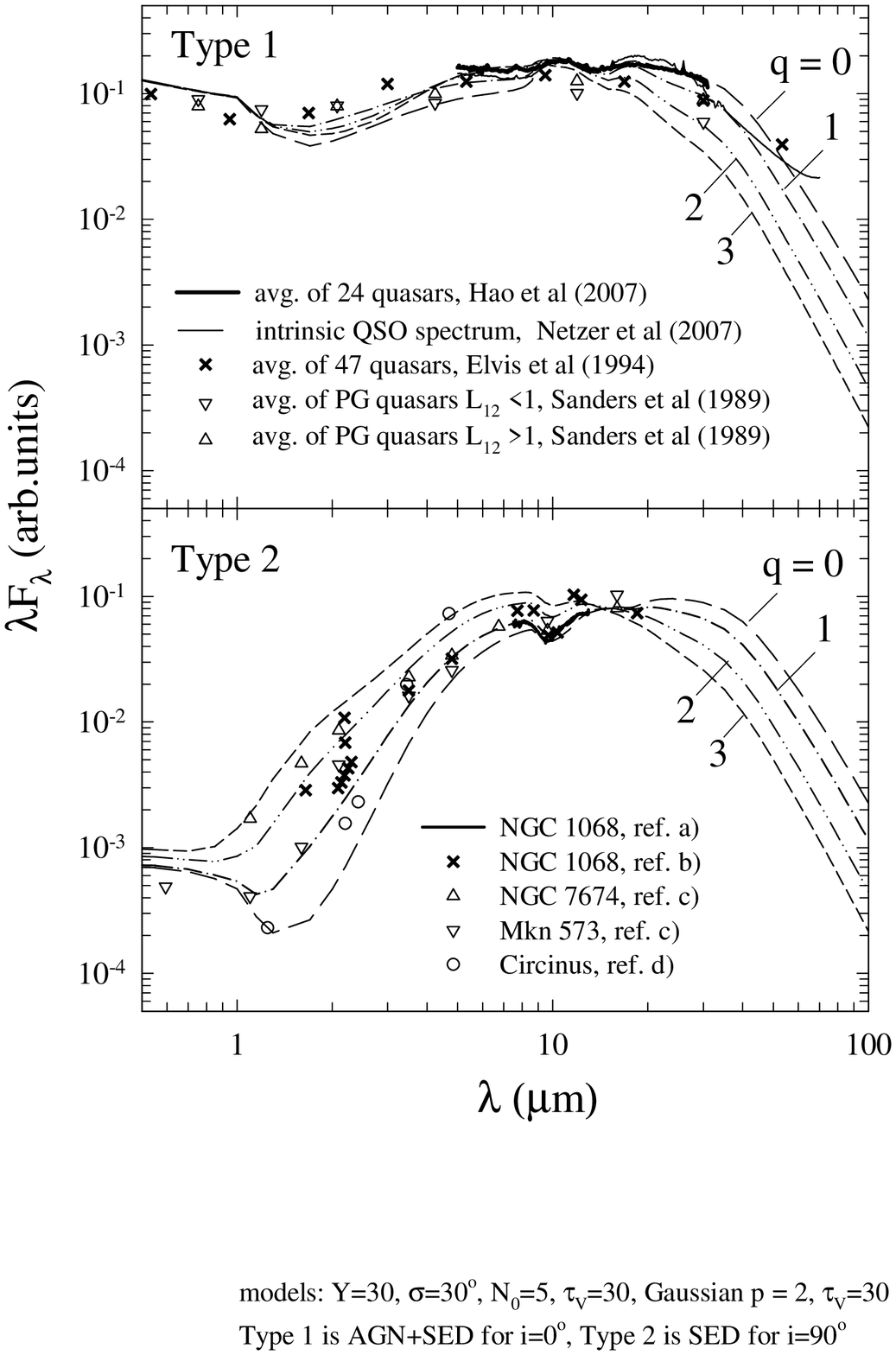}{\figsize}

\caption{Observations of type 1 and type 2 sources compared with clumpy torus
model spectra. The type 1 composite data are from \cite{Sanders89},
\cite{Elvis94}, \cite{Hao07} and \cite{Netzer07}. The type 2 data are from the
following sources: a) \cite{Mason06}; b) various observations with aperture
$\le$ 0.5\arcsec\ listed in \cite{Mason06}; c) \cite{Almudena03}; d)
\cite{Prieto+04}. In the model calculations, plotted with broken lines, each
cloud has optical depth \tV\ = 30. Other parameters are $\sigma$ = 30\deg, $q$
= 0--3, as marked, $Y$ = 30 and \No\ = 5. The angular distribution in this and
all subsequent figures is Gaussian. The models in the upper panel are for
pole-on viewing ($i$ = 0\deg), in the bottom panel for edge-on ($i$ = 90\deg).
} \label{Fig:DataComp}
\end{figure}
%%%%%%%%%%%%%%%%%%%%%%%%%%%%%%%%%%%%%%%%%%%%%%%%%%%%%%%
\fi

\section{MODEL SPECTRA}
\label{sec:SEDs}

We proceed now with the model results. In all calculations the AGN input
radiation follows the ``standard'' spectrum described in \S3.1.1, part I.

\subsection{Geometrical Shape}
\label{sec:geometry}

Figure \ref{Fig:srpGauss} shows model results for sharp-edged and Gaussian
angular distributions. The sharp-edge geometry produces a bimodal distribution
of spectral shapes, with little dependence on viewing angle other than the
abrupt change that occurs between the torus opening and the obscured region. In
contrast, the Gaussian distribution produces a larger variety in model spectral
shapes, with a smooth, continuous dependence on $i$. We investigated a larger
family of angular distributions of the form $\NT(\beta) = \No\exp(-
\left|{\beta/\sigma}\right|^m)$, with $m$ a free parameter. In this family, $m
= 2$ is the Gaussian and as $m$ increases, the transition region around $\beta
= \sigma$ becomes steeper. Generally, ``softer'' distributions with $m \la 10$
show behavior similar to the Gaussian while those with larger $m$ produce
results similar to the sharp-edge geometry.

The SED dichotomy produced by sharp boundaries conflicts with observations.
\cite{Almudena03} studied the 0.4--16 \mic\ nuclear emission from a complete
sample of 58 Seyfert galaxies, selected from the CfA sample. In a comparison
with theoretical models, \citeauthor{Almudena03} point out that a common
prediction of all smooth-density models is a dichotomy of SED between type 1
and 2, similar to the one displayed in the upper panel of fig.\
\ref{Fig:srpGauss}, and that such a dichotomy is not observed in their sample;
the dichotomy is present even in model geometries with soft edges because the
$\exp(-\tau)$ attenuation factor varies rapidly, resulting in a sharp
transition around $\tau \sim 1$ between dusty and dust-free viewing. As is
evident from the lower panel of fig.~\ref{Fig:srpGauss}, this SED dichotomy
problem is solved by soft-edge clumpy tori. Therefore, in the following we
consider only Gaussian angular distributions.

\subsection{Observations and Model Parameters}

As discussed in the Introduction, torus IR observations are hampered by
uncertainties that are partially alleviated by considering composite spectra.
Figure \ref{Fig:DataComp} shows compilations of type 1 and type 2 data and some
representative models, updating a similar figure presented in \cite{NIE02}. The
type 1 data additionally include the recent \emph{Spitzer} composite spectra
from \cite{Hao07} and \cite{Netzer07}. The close agreement between these two
SEDs in their common spectral region, $\lambda$ = 5--38\mic, indicates that
they may have captured the torus emission in outline, if not in details.  The
upturn around 60\mic\ in the Netzer et al spectrum likely reflects the
transition to starburst dominance. To ensure the smallest possible apertures in
type 2 sources, the data for individual objects are mostly limited to ground
based and HST observations.  The data in both panels of this figure display the
general characteristics that have to be reproduced by the same models in
pole-on and edge-on viewing. The updated models plotted with the data differ
from the original ones in \cite{NIE02} in three significant ways: (1) the
optical properties of the silicate component of the dust are taken from the
tabulation for ``cool'' silicates in \cite{OHM} instead of the \cite{Draine84}
dust; (2) the clouds angular distribution is Gaussian rather than sharp-edged;
and (3) the torus radial thickness $Y$ is 30 instead of 100. As is evident from
the figure, the model spectra are generally in reasonable agreement with the
data.

We produced a large number of models for various parameter
sets\footnote{Tabulations of all the models discussed here as well as many
additional cases are available at \url{http://www.pa.uky.edu/clumpy/}}, and we
now present model results and discuss their observational implications. The
models are characterized by free parameters that describe individual clouds
(\tV), control the total number of clouds (\No) and specify the geometrical
properties of their angular and radial spatial distributions ($\sigma$, $q$ and
$Y$). Note that, except for \No, smooth-density models also require all of
these parameters to describe the dust distribution. In the following, the
parameters are varied one at a time, and from comparison with observations we
attempt to identify the likely range of each of them. The effect of the radial
thickness parameter $Y$ is described separately in the next section
(\S\ref{sec:size}), devoted to a discussion of the torus size.

\ifemulate
%%%%%%%%%%%%%%% tauV dependence %%%%%%%%%%%%%%
\begin{figure}
 \Figure{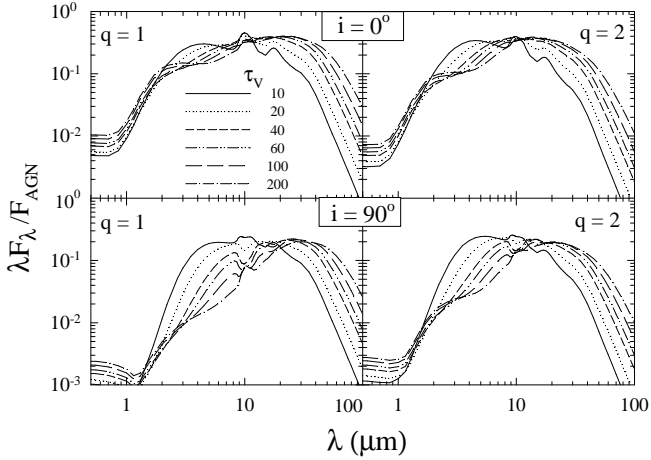}{\figsize}
\caption{Dependence of the torus SED on the single cloud optical depth \tV.
Other parameters are $\sigma = 45\deg$, \No\ = 5 and $Y$ = 30. Radial power law
with $q$ = 1 in the left panels, $q$ = 2 in the right ones. Pole-on viewing in
the top panels, edge-on at the bottom.
} \label{Fig:taudep}
\end{figure}
%%%%%%%%%%%%%%%%%%%%%%%%%%%%%%%%%%%%%%%%%%%%%%%%%%%%%%%
\fi

%\newpage
\subsection{Single Cloud Optical Depth}

Figure \ref{Fig:taudep} shows the effect of varying the optical depth of
individual clouds from 10 to 200. The SED hardly varies when \tV\ increases
beyond \about\ 100, reflecting the similar behavior for emission from a single
cloud (see Fig.\ 11, Part I). The figure shows the torus emission for both
pole-on and edge-on viewing. Smooth-density models \citep[e.g.,][]{PK92,
Granato94} consistently produce the 10 \mic\ silicate feature in emission and
absorption, respectively, for polar and equatorial viewing. As the figure
shows, in a clumpy distribution the feature displays a more complex pattern,
unlike anything produced in smooth-density models. At $i$ = 0\deg, the feature
appears in emission as long as $\tV \la$ 20. When the optical depth increases
further, the feature disappears and the SED is essentially featureless across
the 10\mic\ region. However, the feature reappears in weak emission when $\tV
\ga$ 100. At $i$ = 90\deg, a weak, broad emission feature is evident when \tV\
= 10. When $\tV \ge$ 20, the spectra display a clear absorption feature;
although similar to that of smooth-density models, the feature is never deep,
reflecting the shallow absorption displayed by a single cloud (see \S4.5 part
I). A most peculiar result is the reversal from absorption to an emission
feature, which emerges when \tV\ increases beyond \about\ 100.

The complex behavior of the 10 \mic\ feature arises from a rather intricate
interplay between the emission spectrum of a single cloud and the collective
effect of the entire cloud ensemble. The different patterns can be understood
in terms of the competition between emission and absorption along a given path,
taking account of the flattening of the escape factor \Pesc\ across the
10\,\mic\ feature when \tV\ is increasing (fig.\ \ref{Fig:Pesc}). The behavior
of the 10 \mic\ feature is studied separately at greater depth in
\S\ref{sec:sil} below.

\ifemulate
%%%%%%%%%%%%%%% Torus emission - NT dependence %%%%%%%%%%%%%%
\begin{figure}
 \Figure{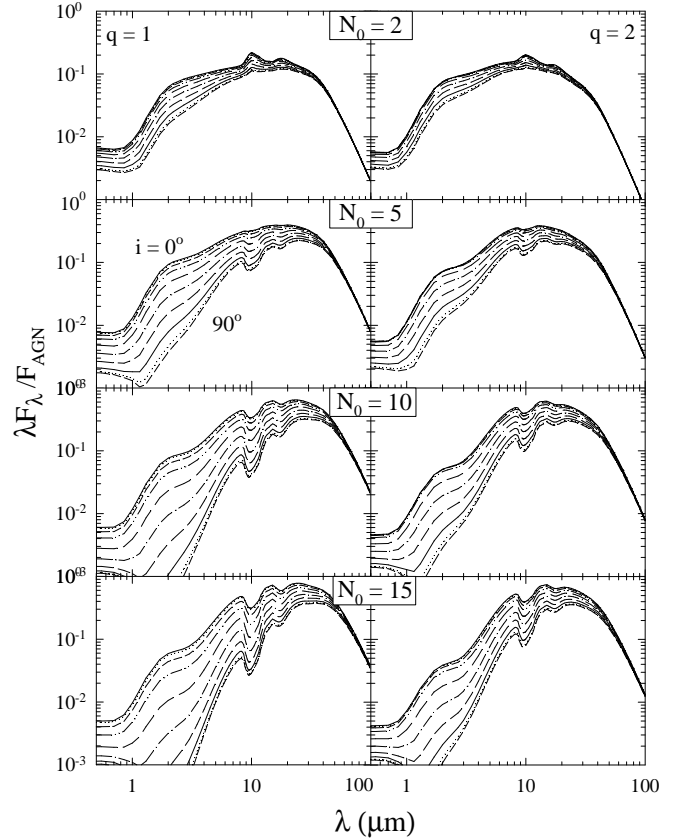}{\figsize}
\caption{Dependence of the torus SED on the number of clouds along a radial
equatorial ray. Each cloud has \tV\ = 60. Angular width $\sigma$ = 45\deg,
power law radial distribution with $q$ = 1 (left panels) and 2 (right),
extending to $Y$ = 30. Different curves in each panel show viewing angles that
vary from 0\deg\ (top curve) to 90\deg\ (bottom) in 10\deg\ steps.
}
 \label{Fig:SED-torus}
\end{figure}
%%%%%%%%%%%%%%%%%%%%%%%%%%%%%%%%%%%%%%%%%%%%%%%%%%%%%%%

%%%%%%%%%%%%%%% NT dependence with added AGN when prob.>50% %%%%%%%%%%%%%%
\begin{figure}
 \Figure{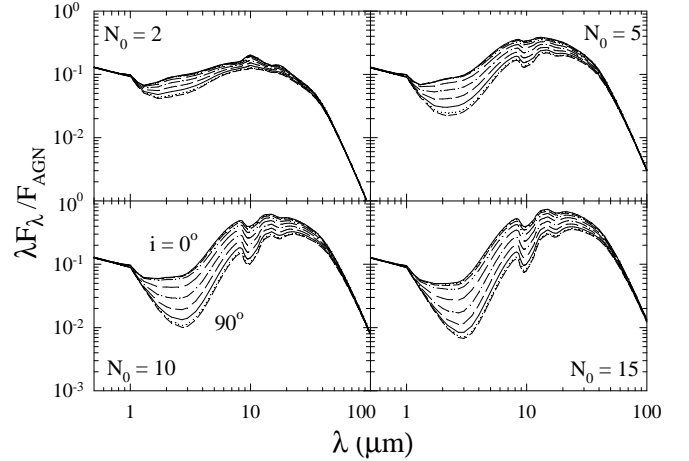}{\figsize}

\caption{Model spectra as in figure \ref{Fig:SED-torus} for $q$ = 2, only with
the AGN contribution added. For each set of parameters, the probability that
the AGN emission will actually be observed can be read from the plots of \Pesc\
in the lower panel of figure \ref{Fig:Pesc}. The break in the SED at $\lambda$
= 1 \mic\ is an artefact of our parametrization of the input spectrum (see
\S3.1.1, part I).} \label{Fig:SED}
\end{figure}
%%%%%%%%%%%%%%%%%%%%%%%%%%%%%%%%%%%%%%%%%%%%%%%%%%%%%%%
\fi

%\newpage

\subsection{Number of Clouds}

Figure \ref{Fig:SED-torus} shows model spectra of torus emission when \No, the
average of the total number of clouds along radial equatorial rays, varies from
2 to 15. The models produce broad IR emission in the $\lambda$ \about\ 1--100
\mic\ range. Values of \No\ larger than 15 produce a very
narrow IR bump peaking beyond 60~\mic. Such SED's have not been observed thus
far, therefore \No\ is likely no larger than \about 10--15 at most.

As is evident from figure \ref{Fig:SED-torus}, when \No\ increases, the
emission in the near- to mid-IR region steepens considerably for viewing close
to equatorial. Composite IR SED of Seyfert 2 galaxies constructed by
\cite{Silva04} show only mild dependence on X-ray absorbing column density as long
as $N_{\rm H} \le$ \E{24} \cs, with considerable steepening when $N_{\rm H}$ is
in the range \E{24}--\E{25} \cs. This behavior is similar to the \No-dependence
displayed in fig.\ \ref{Fig:SED-torus}, thus Compton thick X-ray absorbing
columns might be correlated with a larger \No. For pole-on viewing, the 10
\mic\ feature appears in weak emission when $\No < 5$. As \No\ increases, the
emission switches to absorption that deepens with \No. Moving away from the
axis, the feature displays weak emission when \No\ = 2 but appears in
absorption in all other cases.

In contrast with the smooth-density case, clumpy models always display some
emission at $\lambda < 1\,\mic$ that arises from scattering of the AGN
radiation toward the observer by clouds on the torus far side. Some fraction of
this radiation will always get through the torus near side. The probability for
that is controlled purely by the number of clouds since individual clouds are
always optically thick at UV and optical wavelengths. Varying the number of
clouds produces two competing effects, most clearly visible in figure
\ref{Fig:SED-torus} from the behavior of the $q$ = 1 models at $i$ = 0\deg.
Increasing the number of clouds from \No\ = 2 to 5 raises the level of the
radiation that gets through because there are more scattering clouds. With
further increase in \No, obscuration by intervening clouds takes over and the
emerging intensity decreases. It is hard to assess the observational
significance of this aspect of the results. Our models include a single type of
clouds and no intercloud medium. Such a medium with an optical depth \tV\ of
only a few would attenuate all wavelengths shorter than \about 1\,\mic\ in the
model spectra without significantly affecting the infrared. We plan a detailed
study of these effects in future work.

Like most models presented here, fig.\ \ref{Fig:SED-torus} shows only the torus
emission, corresponding to type 2 SED. Our model predictions for type 1 SED can
always be obtained by adding the AGN direct radiation. However, unlike the
smooth-density case, the probability for a clear view of the AGN depends not
only on the viewing angle but also on the number of clouds (see
\S\ref{sec:the_AGN}). Figure \ref{Fig:SED} displays again the $q$ = 2 models
shown in fig.\ \ref{Fig:SED-torus}, but this time the AGN contribution is added
in. The probability that this would be the SED actually detected in a given
source is given by the corresponding \Pesc, shown in the lower panel of figure
\ref{Fig:Pesc}. When visible, the AGN dominates the emission at $\lambda \la$
3\mic. The transition from AGN to torus domination of the SED is an important
issue that requires detailed observations of type 1 sources in the near- and
mid-IR regions.

\ifemulate
%%%%%%%%%%%%%%%%%%%%%%%%%%%%%%%%%%%%%%%%%%%%%%%%%%%%%%%
\begin{figure}
 \Figure{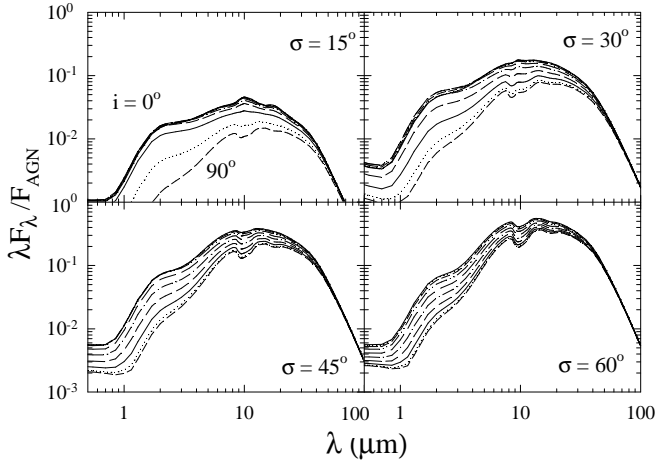}{\figsize}
\caption{Dependence of the torus model spectra on the width
parameter $\sigma$ of the Gaussian angular distribution. Each cloud
has \tV\ = 60. The cloud radial distribution is a power law with
\No\ = 5 and $q$ = 2 extending to $Y$ = 30. Viewing angles vary from
0\deg\ to 90\deg\ in 10\deg\ steps. } \label{Fig:SED-opang}
\end{figure}
%%%%%%%%%%%%%%%%%%%%%%%%%%%%%%%%%%%%%%%%%%%%%%%%%%%%%%%
\fi

\subsection{The Torus Angular Width}

The effect of the angular distribution width is shown in figure
\ref{Fig:SED-opang}, which displays results for a few representative $\sigma$.
The spectral shapes of models with $\sigma = 15\deg$ are in general agreement
with observed SED but the dependence on viewing angle displays a bi-modal
distribution that conflicts with observations of Seyfert galaxies
(\S\ref{sec:geometry}). Values in the range 30\deg--50\deg\ produce similar
spectral shapes, all in general agreement with observations.  The $\sigma =
30\deg$ models provide the best match to the behavior of the 10\mic\ feature in
the average spectra of Seyfert 1 and 2 galaxies (see \S\ref{sec:sil}).
Estimates of the torus angular width based on statistics of Seyfert galaxies
that take proper account of clumpiness give $\sigma$ \about\ 30\deg\ (see
\S\ref{sec:unification}). At $\sigma$ = 60\deg, the 10\mic\ feature appears in
pronounced absorption at all viewing angles. Increasing the width parameter
further all the way to $\sigma$ = 85\deg\ has little effect on the SED, except
that the dependence on viewing angle decreases, as is expected from the
approach to spherical symmetry.

\ifemulate
%%%%%%%%%%%%%%%%%%%%%%%%%%%%%%%%%%%%%%%%%%%%%%%%%%%%%%
\begin{figure}
 \Figure{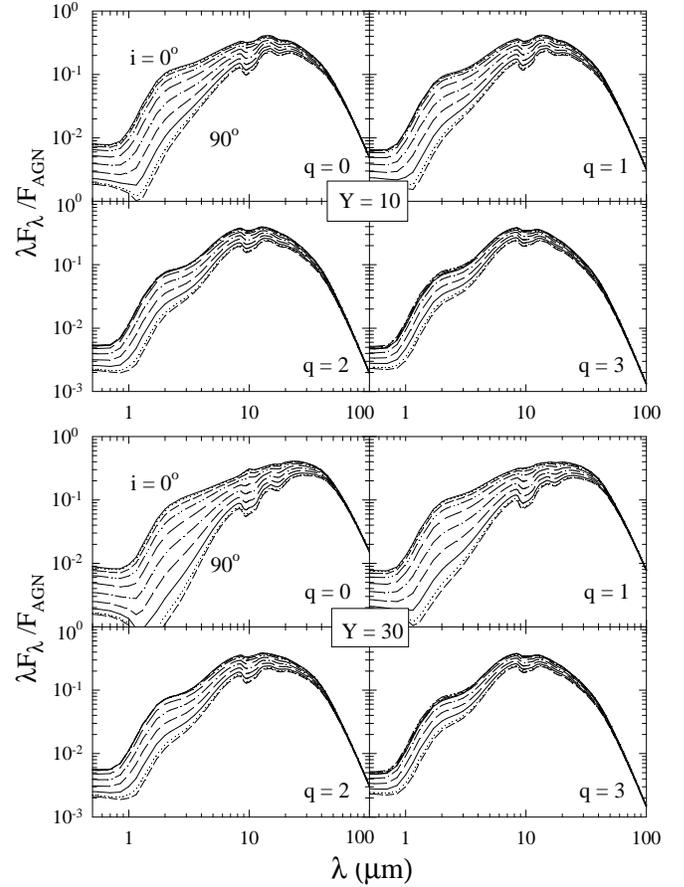}{\figsize}

\caption{Dependence of the torus model spectra on the power $q$ of the radial
density distribution, which extends to $Y$ = 10 in the top panels and $Y$ = 30
in the bottom ones. \No\ = 5 clouds with \tV\ = 60 each. Angular width $\sigma$
= 45\deg. Viewing angles from 0\deg\ to 90\deg\ in 10\deg\ steps. The emission
anisotropy decreases when $q$ increases.} \label{Fig:SED-qdep}
\end{figure}
%%%%%%%%%%%%%%%%%%%%%%%%%%%%%%%%%%%%%%%%%%%%%%%%%%%%%%%
\fi

%\newpage

\subsection{Radial Profile and IR Emission Anisotropy}
\label{sec:Radial}

Figure \ref{Fig:SED-qdep} shows the SED when $q$, the index of the power-law
radial distribution, varies in the range 0--3 for two values of the radial
thickness $Y$. Since \No\ is kept fixed, varying $q$ changes only the placement
of clouds between the inner, hotter parts and the outer, cooler regions,
shifting the emission between near- and far-IR. Steep radial distributions ($q$
= 2, 3) produce nearly identical spectral shapes for $Y$ = 10 and 30 because
the clouds are concentrated near the inner boundary in these cases.

The variation of SED with viewing angle displayed in fig.\ \ref{Fig:SED-qdep}
is much smaller than in smooth-density models \citep{PK92, PK93, EfRR95,
Granato94, Granato97, Dullemond05, Schartmann05}. For example, our models
produce at 1 \mic\ an edge-to-pole flux ratio of about 5 or less for cloud
optical depth of $\tau_V=60$ . In Fig. 1 from \citeauthor{Granato97}, the
corresponding flux ratio is several hundred for $A_V=30$, and off-scale for
$A_V=100$. The degree of isotropy is especially high when the torus is small,
but even in the case of $Y$ = 30 the emission becomes nearly isotropic as $q$
increases (the radial distribution gets steeper). This point is further
illustrated in Figure \ref{Fig:Aniso-lam}, which shows the viewing-angle
variation of the observed flux at different wavelengths. Since the torus flux
is normalized to the AGN overall flux, the quantity plotted in this figure
provides the bolometric correction for each of the displayed wavelengths. The
anisotropy decreases with wavelength, practically disappearing beyond the
mid-IR --- at $\lambda$ = 12 \mic\ the variation with viewing-angle is within a
factor of \about 2 for both $q$ = 1 and 2. Another indicator of the emission
anisotropy is the variation of the torus bolometric flux \Ftor\ = $\int
F_\lambda d\lambda$ with viewing angle, shown in figure \ref{Fig:fbol} for $q$
= 2. The variation increases with the number of clouds but remains less than a
factor of 3 even when \No\ = 15, the likely upper limit.

\ifemulate
%%%%%%%%%%%%%%%%%%%%%%%%%%%%%%%%%%%%%%%%%%%%%%%%%%%%%%%
\begin{figure}
 \Figure{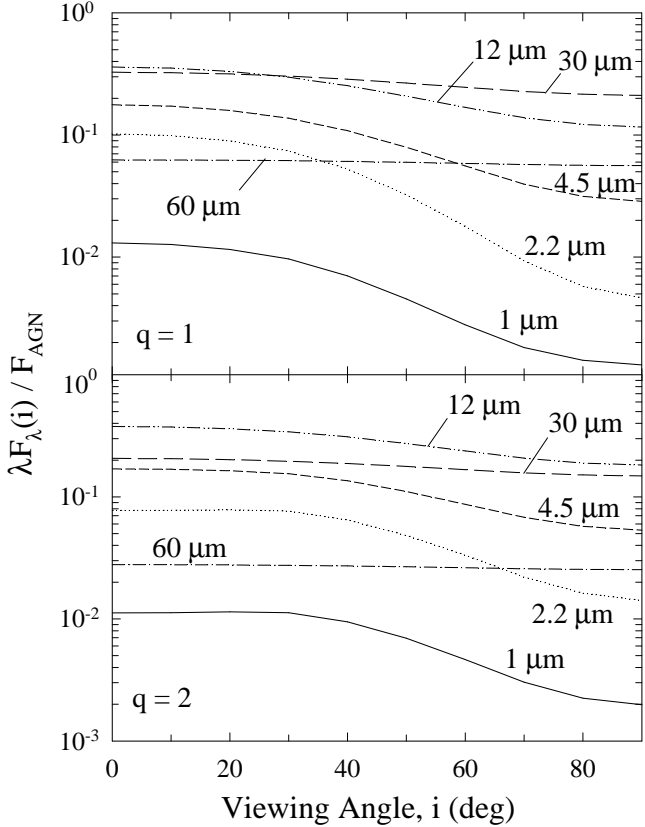}{\figsize}

\caption{Variation of the torus flux with viewing angle at different
wavelengths, as marked. \No\ = 5 clouds with \tV\ = 60 each in a radial
distribution with $q$ = 1 (top panel) and 2 (bottom) extending to $Y$ = 30.
Angular width $\sigma$ = 45\deg. Since the emission is normalized to the AGN
flux, the plotted quantity provides the bolometric correction for each
displayed wavelength.} \label{Fig:Aniso-lam}
\end{figure}
%%%%%%%%%%%%%%%%%%%%%%%%%%%%%%%%%%%%%%%%%%%%%%%%%%%%%%%

%%%%%%%%%%%%%%%%%%%%%%%%%%%%%%%%%%%%%%%%%%%%%%%%%%%%%%%
\begin{figure}
 \Figure{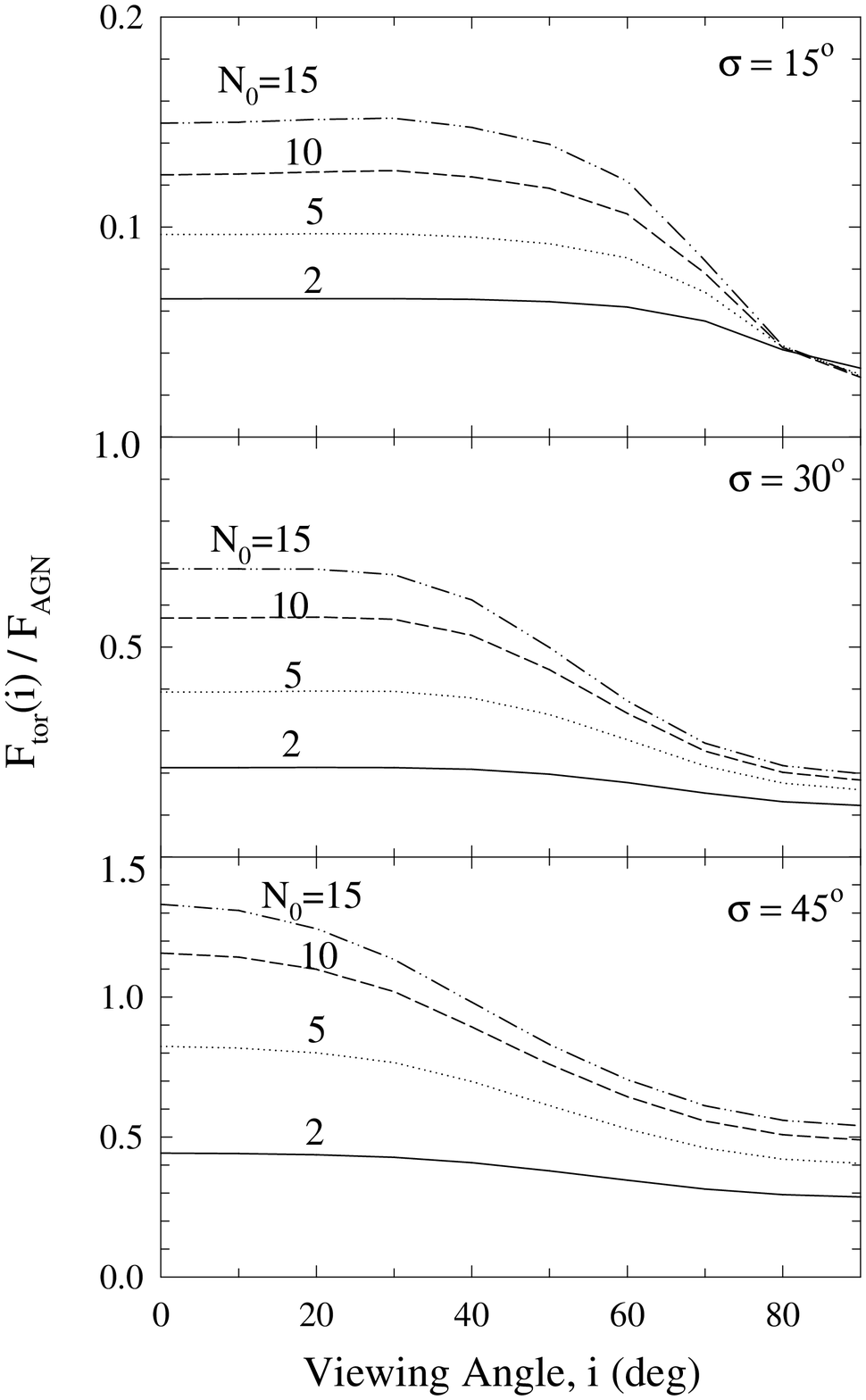}{0.85\figsize}
 \Figure{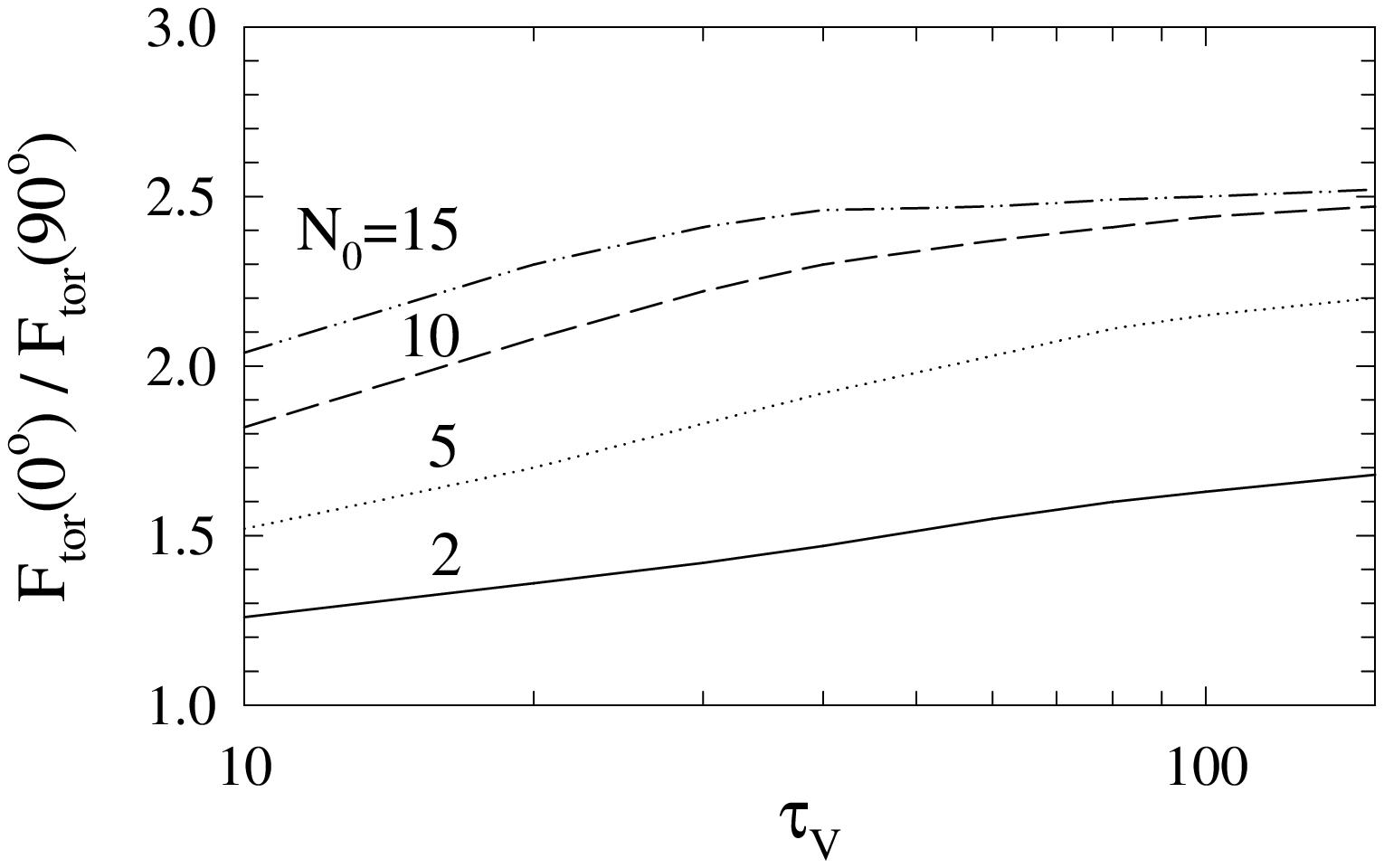}{0.85\figsize}
\caption{Anisotropy indicators for the torus bolometric flux \Ftor. All models
have $q$ = 2 and $Y$ = 30. {\em Top:} Variation of \Ftor\ with viewing angle
when \tV\ = 60 for various values of \No\ (cloud number) and $\sigma$
(torus angular width), as indicated; note the changing vertical scale. {\em
Bottom:} Ratio of the torus bolometric fluxes along the axis and equator as a
function of single cloud optical depth. The torus angular width is $\sigma$ =
45\degr.}
 \label{Fig:fbol}
\end{figure}
%%%%%%%%%%%%%%%%%%%%%%%%%%%%%%%%%%%%%%%%%%%%%%%%%%%%%%%
\fi

It is important to note that at every viewing angle, the AGN obscuration is
identical in all the models displayed in fig.\ \ref{Fig:SED-qdep}; obscuration
depends only on the total number of clouds along radial rays, which is the same
in all cases. Indeed, for each of these models the probability for direct view
of the AGN as a function of $i$ is shown by the \No\ = 5 curve in the bottom
panel of figure \ref{Fig:Pesc}. Along this curve, \Pesc\ varies by two orders
of magnitudes between polar and equatorial viewing. That is, a clumpy torus can
produce extremely anisotropic obscuration of the AGN {\em together with nearly
isotropic mid-IR emission}.

Recent observations seem to indicate that this is indeed the required behavior.
Ground based observations of AGN nuclear emission at 10\mic\ show it to be well
correlated with the hard X-ray luminosity, and both type 1 and type 2 sources
follow the same correlation \citep{Almudena01, Krabbe01}. \cite{Whys04} report
that the mean values for the 12\,\mic/60\,\mic\ flux ratios of Seyfert 1 and 2
galaxies differ by only \about 30\%. Since the 60 \mic\ emission is optically
thin and thus essentially isotropic (cf fig.\ \ref{Fig:Aniso-lam}), this result
indicates that the variation of the 12 \mic\ flux is small. \cite{Lutz04}
compared the 6 \mic\ ISO fluxes of Seyfert 1 and Seyfert 2 galaxies normalized
to their intrinsic hard X-ray fluxes. They conclude that the distributions of
the two populations are essentially identical within the observational errors,
and note the conflict with the anisotropy predicted by smooth-density torus
models. The \citeauthor{Lutz04}\ finding was confirmed by \cite{Horst06}, who
used the same approach with ground based, and thus better resolution,
observations at 12 \mic. \cite{Buchanan06} conducted Spitzer observations of 87
Seyfert galaxies in the $\lambda$ = 5--35 \mic\ range and normalized the IR
fluxes with the optically thin radio emission. Although at 6 \mic\ they find a
larger variation than \citeauthor{Lutz04}, they also find that the emission
from Seyfert 1 and 2 galaxies are within factor 2 of each other for all
$\lambda \ga$ 15 \mic, and note the discrepancy with smooth-density models.
Finally, the average spectra of Seyfert 1 and 2 galaxies derived by
\cite{Hao07} from Spitzer observations have nearly identical shapes, except for
the 10\mic\ silicate region.

The moderate level of anisotropy found in the observations suggests that, if
the torus radial thickness is $\ga$ 20 then the steeper radial profile $q$ = 2
might be more appropriate than $q$ = 1. It may be noted that the clumpy torus
models in \cite{Mason06}, which utilized $Y$ = 100, yielded the best fits to
the observations of NGC 1068 with $q$ = 2.

%\newpage
\section{TORUS SIZE}
\label{sec:size}

The fraction of the sky obscured by the torus determines the relative numbers
of type 1 and 2 sources, and the statistics of Seyfert galaxies show that the
height and radius of obscuring dusty torus obey $H/R$ \about\ 1 (see \S\ref{sec:unification}).
Since obscuration does not depend separately on either $H$ or $R$, only on
their ratio, neither quantity is determined individually. An actual size can
only be determined from the torus emission.

\subsection{SED Analysis}

\ifemulate
%%%%%%%%%%%%%%%%%%%%%%%%%%%%%%%%%%%%%%%%%%%%%%%%%%%%%%
\begin{figure}
 \Figure{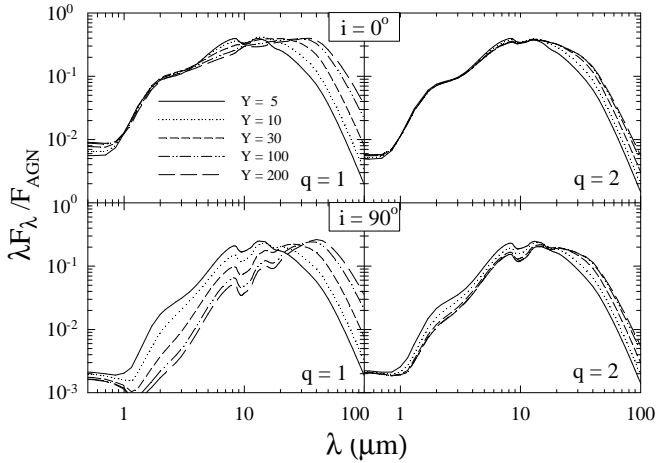}{\figsize}

\caption{Dependence of the SED of a clumpy torus on the radial thickness $Y =
\Ro/\Rd$, as marked. Radial distribution with $q$ = 1 (left panels) and 2
(right). All models have \No\ = 5 clouds with \tV\ = 60 each, and $\sigma$ =
45\deg. Pole-on viewing in the top panels, edge-on at the bottom. Note that the
curves in the left-bottom panel have a similar shape at $\lambda \la 15$ \mic\
and would nearly overlap if normalized to a common wavelength in that range
instead of \FAGN. }
 \label{Fig:SED-Ydep}
\end{figure}
%%%%%%%%%%%%%%%%%%%%%%%%%%%%%%%%%%%%%%%%%%%%%%%%%%%%%%

%%%%%%%%%%%%%%%%%%%%%%%%%%%%%%%%%%%%%%%%%%%%%%%%%%%%%%
\begin{figure*}[ht]
\begin{minipage}{\textwidth}
 \Figure{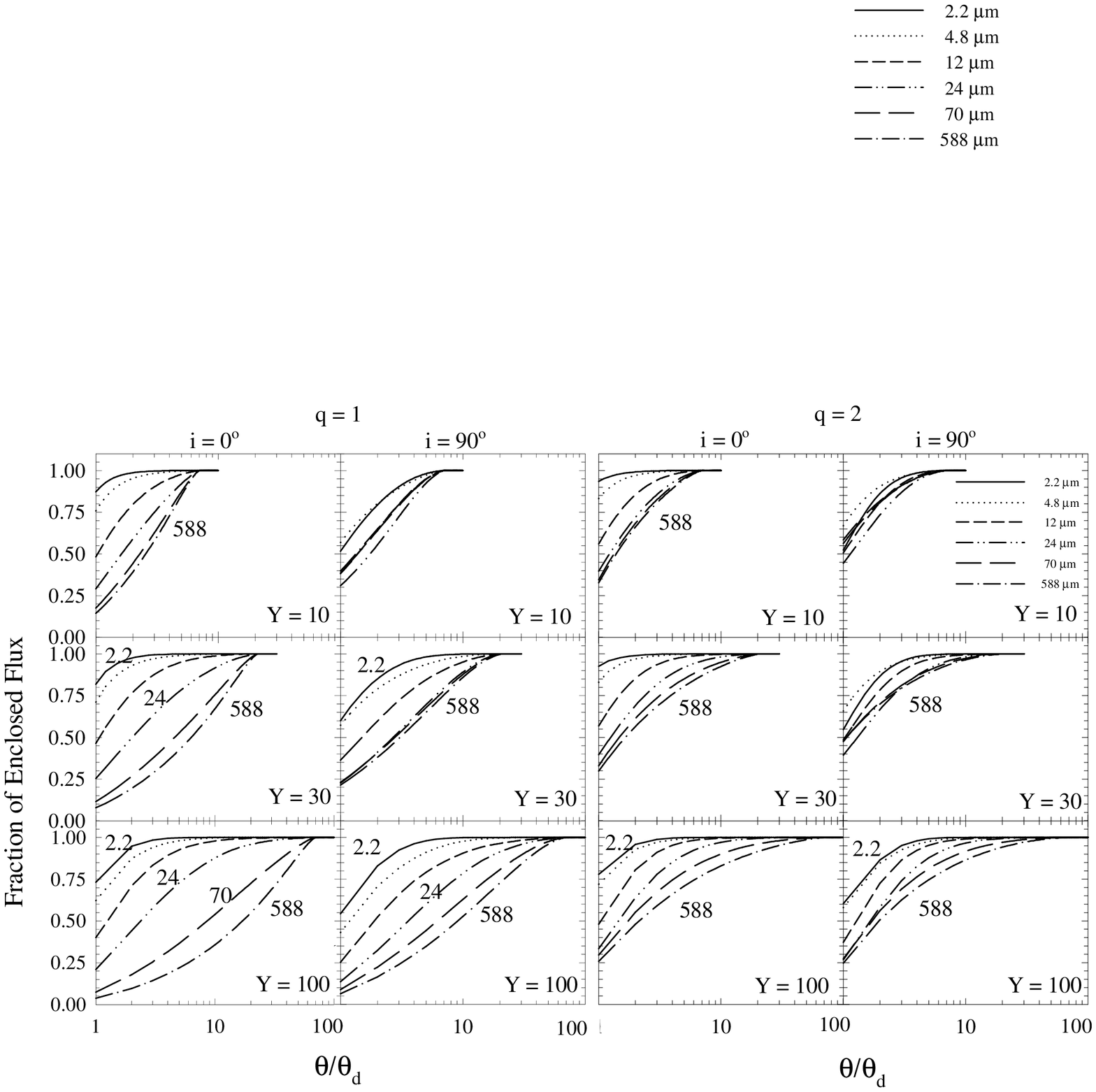}{0.9\hsize}
\caption{Fraction of the torus flux enclosed within a circle with angular
radius $\theta$ centered on the AGN. Wavelengths, in \mic, as labeled (some
labels omitted for clarity). \td\ is the angle equivalent of the torus inner
radius \Rd\ (eq.\ \ref{eq:Rd}) at the observer's location; for reference, \td =
0\farcs02 for \Rd\ = 1\,pc at 10 Mpc. All models have \No\ = 5, \tV\ = 60,
$\sigma = 45\deg$ and $q$ = 1 or $q$ = 2, as marked. In each case, pole-on
viewing in the left panels, edge-on at right. Torus sizes, from top to bottom,
are $Y$ = 10, 30 and 100. }
 \label{Fig:FracFlx}
\end{minipage}
\end{figure*}
%%%%%%%%%%%%%%%%%%%%%%%%%%%%%%%%%%%%%%%%%%%%%%%%%%%%%%%%%%%%%%%%%
\fi

In the absence of high-resolution IR observations, early estimates of the torus
size came from theoretical analysis of the SED. For a given dust sublimation
temperature, the torus inner radius \Rd\ is determined from the AGN luminosity
(eq.\ \ref{eq:Rd}). The dust temperature distribution, and with it all model
results, depends only on $r/\Rd$. Therefore, the only size parameter that can
be determined from SED modeling is the radial thickness $Y = \Ro/\Rd$.
\cite{PK92} performed the first detailed calculations with a uniform density
torus and found $Y$ \about\ 5--10. However, in subsequent work \cite{PK93}
speculated that this compact structure might be embedded in a much larger, and
more diffuse torus, extending typically to \about\ 30--100 pc. \cite{Granato94}
extended the smooth-density calculations to more elaborate toroidal geometries.
They conclude that ``The broadness of the IR continuum of Seyfert 1 nuclei
requires an almost homogeneous dust distribution extending at least to a few
hundred pc (\Ro/\Rd\ $\ga$ 300 or \Ro\ $\ga$ 300$L_{46}^{1/2}$ pc)'' and that
``broad (\Ro\ $\simeq$ 1000$L_{46}^{1/2}$ pc) tori'' would be ``fully
consistent with available broad-band data and high-resolution IR spectra of
Seyfert 1 and 2 nuclei''. Although subsequent modeling produced somewhat
smaller sizes \citep{Granato97, Fritz06}, the original requirements of uniform
density and large dimensions directly reflect the large amounts of cool dust
necessary for producing the torus IR emission. This requirement arises because
in smooth density distributions, the dust temperature is uniquely related to
distance from the AGN. While this statement is strictly correct only for
single-size dust grains, even when dust size distribution in invoked in
smooth dust models, the observations still favor clumpy dust distribution
\citep{Schartmann05}.

The one-to-one correspondence between distance and temperature does not hold in
clumpy media, where different dust temperatures coexist at the same distance
and where the same temperature can be found at different distances (see part I,
\S3.1.2 and \S4.2). For example, in the model discussed in Fig.~7 from part I,
the dust temperature at $Y=10$ ranges from 150 K to 600 K, while
\cite{Schartmann05} find using smooth models with realistic dust size distribution
that the temperature range at $Y=10$ is 250-300 K (i.e. a ratio of 1.2
vs. 4 for clumpy dust). In contrast with smooth density distributions, a
clumpy torus contains cool dust on the dark sides of clouds much closer to the
heating source and thus can emit IR efficiently from its inner regions.

Figure \ref{Fig:SED-Ydep} shows our model results for the SED of clumpy tori with
various radial thicknesses for $q$ = 1 and 2. In spite of the factor 40
variation in torus thickness, the SED is quite similar for all the $q$ = 2
models. The reason is simple---irrespective of the torus size, at least 80\% of
all clouds are located within $r \le 5\Rd$ in this case. The models with $q$ =
1 display discernible variations, but these variations are mostly confined to
$\lambda \ga 15\mic$. The large differences apparent at shorter wavelengths in
the edge-on viewing do not reflect intrinsic variation of the SED, only the
scaling with the bolometric flux. If these curves were scaled instead to the
same value at, say, 2\mic, they would all overlap up to $\lambda$ \about\
15\mic, similar to the pole-on viewing. These differences can be further
understood with the aid of figure \ref{Fig:FracFlx}, which shows the fraction
of the overall flux contained within circular apertures of increasing size.
This fraction, as well as the brightness distribution, is a function of
$\theta/\td$ (= $r/\Rd$), where $\theta$ is angular displacement from the
center and $\td = \Rd/D$. The figure shows that at wavelengths shorter than
5\,\mic, almost all the flux is originating from inside 3\td\ irrespective of
the value of $Y$. Therefore, such wavelengths cannot determine the torus size.
In the $q$ = 2 case, even longer wavelengths cannot distinguish between the
different sizes because 80\% of the flux always originates from the inner
10\,\td. In the $q$ = 1 case the portions beyond 10\td\ contribute
significantly to the flux of a larger torus, but only at wavelengths longer
than \about 12--15\mic.

These results show that the model SED do display appreciable differences among
tori of different sizes when $q$ = 1 and that determining the torus size in
this case would require measurements of its flux at $\lambda \ga$ 15 \mic.
Because of the large beam sizes at such long wavelengths, current observations
generally cannot distinguish between the contributions of the torus and its
surroundings to the overall flux. For example, as noted in the Introduction, in
NGC 1068 the torus contributes less than 30\% to the 10 \mic\ flux measured
with apertures $\ge$ 1\arcsec\ \citep{Mason06}. The $q$ = 2 density profile,
which might be the more common radial distribution (\S\ref{sec:Radial}), does
not generate any discernible distinctions among spectral shapes. Therefore,
{\em the radial size of a clumpy torus cannot be constrained by SED
measurements}. Only high-resolution observations can determine the torus size
--- the SED does not have the necessary discriminative power.

\ifemulate
%%%%%%%%%%%%%%%%%%%%%%%%%%%%%%%%%%%%%%%%%%%%%%%%%%%%%%%
\begin{figure}
 \Figure{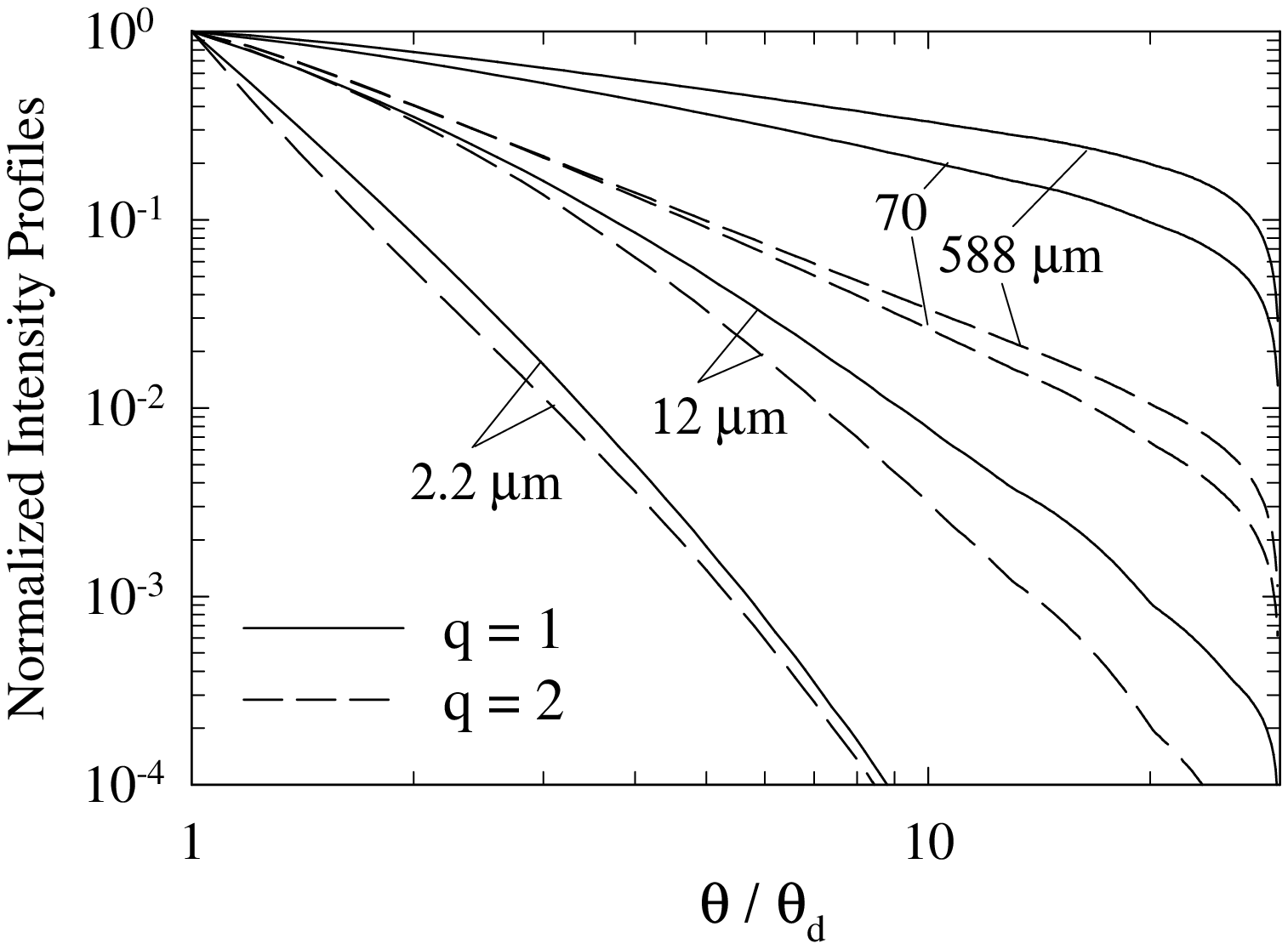}{0.9\figsize} \\
 \Figure{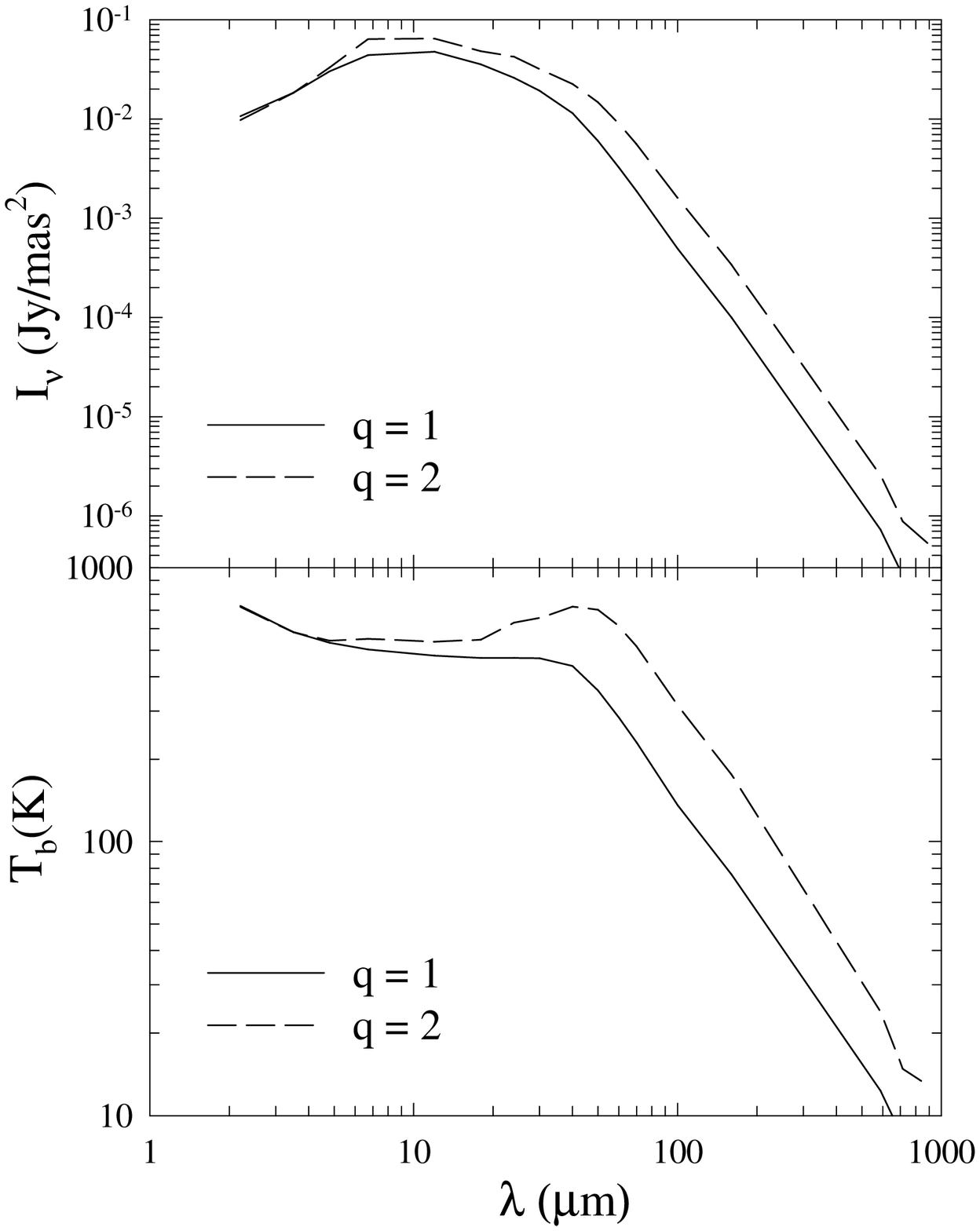}{0.9\figsize}

\caption{Surface brightness for torus models with \tV\ = 60, \No\ = 5, $\sigma$
= 45\deg, $Y$ = 30 and $q$ = 1 and 2, as indicated. The top panel shows the
radial variation of intensity with angular displacement from the center for
pole-on viewing and a set of wavelengths as marked. The AGN emission, which is
not shown, corresponds to a narrow spike at $\theta/\td \ll 1$ (see text). Each
intensity profile is normalized to its brightness level at $\theta = \td$,
shown in the bottom panel together with the corresponding brightness
temperature.
} \label{Fig:IntProfl}
\end{figure}

%%%%%%%%%%%%%%%%%%%%%%%%%%%%%%%%%%%%%%%%%%%%%%%%%%%%%%%
\fi

\subsection{Brightness Profiles}
\label{Sec:BrightProf}

Because of the discrete nature of a clumpy medium, different lines of sight
will generally produce a different brightness even when crossing similar
regions. Our formalism provides the statistical average along such rays;
fluctuations around this average can be large (see \S 2.1, part I).

Figure \ref{Fig:IntProfl} shows model intensity profiles for pole-on viewing of
a torus with Y = 30.  The curves show only the torus emission, starting at its
inner edge $\theta = \td$.  The AGN emission, which is not shown, would produce
a narrow spike between $0 \le \theta \le \theta_{\rm AGN}$, where $\theta_{\rm
AGN}/\td \sim (T_{\rm sub}/T_{\rm b,AGN})^2$ and $T_{\rm b,AGN}$ is the AGN
brightness temperature \citep[e.g.,][]{IE97}. Therefore,  $\theta_{\rm AGN}/\td
\ll 1$ under all circumstances. The set of displayed wavelengths extends from
the $K$-band, where most of the current imaging observations are performed, to
588 \mic, one of the wavelengths that will become available for high-resolution
imaging when ALMA is fully operational. The torus intensity is highest on or
close to its inner edge. The brightness is highest around 12 \mic, as is
evident from the figure bottom panel. For both radial density profiles used in
this figure, the brightness declines to half its peak value within $\theta <
5\,\td$ at all displayed wavelengths. At 12\mic\ and shorter wavelengths, the
brightness declines to 1\% of peak value within $\theta \le 10\,\td$.
Evidently, observations attempting to probe the torus structure must combine
high resolution with a large dynamic range.

Near-IR wavelengths provide little information about the torus structure and
size. As is evident from the figure, at 2 \mic\ it would be difficult to
distinguish between $q$ = 1 and 2 radial density distributions even with
high-resolution observations. The steep brightness decline at these wavelengths
also makes it practically impossible to determine the torus full size. Since
the brightness falls under 1\% of its peak value at $\theta = 4\,\td$ for
either density profile, determining whether the torus ends at that point or
continues to larger radii would be a difficult task. A $Y$ = 10 torus is
indistinguishable from the inner 10\td\ of a torus as large as $Y$ = 100, as is
evident also from figure \ref{Fig:FracFlx}. As the wavelength increases, the
brightness fall-off becomes less steep. VLTI interferometry has angular
resolution of order 0\farcs01 at 12 \mic, but it would still be difficult to
distinguish between the two displayed radial density profiles even in systems
where \td\ is of a similar order of magnitude. The two density profiles produce
distinctly different brightness profiles at 70\mic, but there are no
instruments with the required angular resolution at that wavelength. In the
foreseeable future, ALMA seems to be the only facility with a realistic chance
to determine through 588 \mic\ observations the radial cloud distribution and
whether a torus does extend beyond $Y$ = 30.

\subsection{Observations}

With the advent of high-resolution IR observations, direct imaging is now
available for some AGN tori, and upper limits have been set on the dimensions
of the nuclear IR source in others. Interferometric observations at 8--13 \mic\
with the VLTI have resolved by now the nuclear region in three AGN: NGC 1068,
Circinus and Cen A.  The thermal emission in all three cases is rather compact.
In NGC 1068 \cite{Jaffe04} find that the emission extends to $R$ = 1.7 pc.
\cite{Poncelet06} reanalyzed the same data with slightly different assumptions
and obtained a similar result, $R$ = 2.7 pc. The AGN bolometric luminosity is
\about\ 2\x\E{45} erg\,s$^{-1}$ in this case \citep{Mason06}, so that \Rd\ is
\about 0.6 pc and the torus mid-IR emission is confined within \about
3--5\,\Rd. In Circinus, \cite{Tristram07} find that the torus emission extends
to $R$ = 1 pc.  The AGN bolometric luminosity is \about\ 8\x\E{43}
erg\,s$^{-1}$ \citep{Oliva99}, so \Rd\ $\simeq$ 0.1 pc and the outer radius of
the mid-IR emission is \about 10\Rd. The nature of the mid-IR emission from the
Cen A nucleus is somewhat involved---\cite{Meisenheimer07} conclude that it
contains an unresolved synchrotron core and thermal emission within a radius of
\about 0.3 pc. Since the AGN bolometric luminosity is \about\ 1\x\E{43}
erg\,s$^{-1}$ \citep{Whys04}, \Rd\ is \about 0.04 pc and the torus emission
does not exceed \about 8\Rd\ in this source. One other case of resolved mid-IR
emission involves NGC 7469, where \cite{Soifer03} find a 12.5 \mic\ compact
nuclear structure contained within $R <$ 13 pc. Unfortunately, NGC 7469 is a
clear case where the IR signature is dominated by the starburst component even
though the AGN dominates the optical classification \citep{Weedman05},
therefore the resolved compact structure cannot be identified with the torus
\citep[see also][]{Davies04}.

Although there are no other reports of resolved torus emission at this time,
upper limits on the torus size have been reported in some additional sources.
\cite{Prieto04} studied a number of AGN in the 1--5 \mic\ range. In all cases
the observations show unresolved nuclear emission at these wavelengths, setting
upper limits on the torus radius of $\la$ 5--10 pc, depending on the target
distance. Even more significant are the upper limits reported at mid IR.
\cite{Radomski03} place an upper limit $R < $ 17 pc at 10 \mic\ and 18 \mic\ on
the nuclear component in NGC 4151, while \cite{Soifer03} place the tighter
constraint $R <$ 5 pc at 12.5 \mic. \citeauthor{Soifer03}\ also find an upper
limit $R <$ 14 pc for the 12.5 \mic\ compact nuclear emission in NGC 1275.

\ifemulate
%%%%%%%%%%%%%%%%%%%%%%%%%%%%%%%%%%%%%%%%%%%%%%%%%%%%%%%%%%%%%%%%%%%%%%%%%%%%
\begin{figure*}
\begin{minipage}{\textwidth}
 \centering \leavevmode
 \includegraphics[width=0.8\hsize,clip]{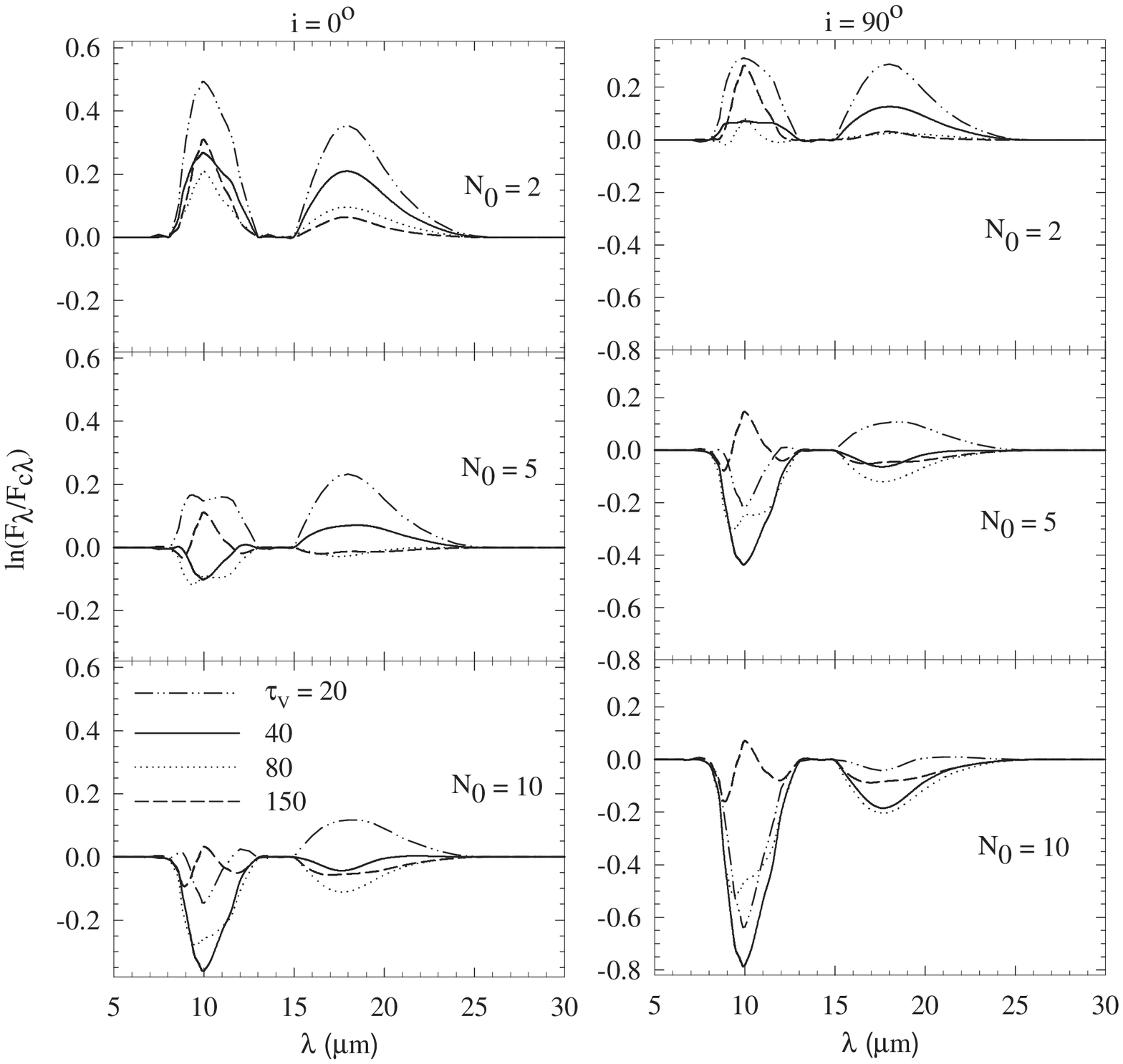}

\caption{Spectral shape of the silicate 10 and 18 \mic\ features: $F_\lambda$
is the torus emission in the 5--30 \mic\ region and \Fcont\ is the smooth
underlying continuum obtained by a spline connecting the feature-free segments
in this spectral region (see text). All models have $q$ = 2, $Y$ = 30, $\sigma$
= 45\deg, and \No\ as marked in each panel. Curves correspond to different \tV,
as labeled. Panels on the left correspond to pole-on viewing, on the right to
edge-on; note the different scales of the vertical axes in the two cases.
} \label{Fig:silfeature}
\end{minipage}
\end{figure*}
%%%%%%%%%%%%%%%%%%%%%%%%%%%%%%%%%%%%%%%%%%%%%%%%%%%%%%%%%%%%%%%%%%%%%%%%%%%%
\fi

\subsection{How Big is the Torus?}
\label{sec:Rout}

All current observations are consistent with a torus radial thickness $Y =
\Ro/\Rd$ that is no more than \about\ 20--30, and perhaps even as small as
\about\ 5--10. Although larger values cannot be ruled out, nothing in the
currently available IR data requires their existence. Similarly, molecular line
observations do not give any evidence for large toroidal structures with the
height-to-radius ratio $H/R$ \about\ 1 required from unification statistics
(see \S\ref{sec:unification} below). In NGC 1068, \cite{Schinnerer00} find from
CO velocity dispersions that at $R \simeq$ 70 pc the height of the molecular
cloud distribution is only $H$ \about\ 9--10 pc, for $H/R$ \about\ 0.15.
\cite{Galliano03} model H$_2$ and CO emission from the same source with a
clumpy molecular disk with radius 140 pc and scale height 20 pc, for the same
$H/R$ \about\ 0.15. Therefore, although resembling the putative torus, the
distribution of these clouds does not meet the unification scheme requirement
$H/R$ \about\ 1. Evidently, the detected molecular clouds are located in a
thinner disk-like structure outside the torus. Recent 10\mic\ imaging
polarimetry of NGC 1068 by \cite{Packham07} shed some light on the continuity
between the torus and the host galaxy's nuclear environments.

As is evident from the above discussion, determining the torus actual endpoint
is rather difficult, if not impossible; in fact, insisting on an endpoint for a
steep $1/r^2$ distribution is meaningless in practice (with the currently
available observations). The torus is embedded in the central region of the
host galaxy, and the steep radial decline of its brightness implies that its
emission is unlikely to be cleanly separated from the surroundings. The only
observations holding realistic chance for doing that are future high resolution
sub-mm measurements with ALMA. Even those would require detailed analysis that
takes into account the emission from both the torus and its surrounding.

It seems safe to conclude that there is no compelling evidence at this time
that torus clouds beyond $Y$ \about\ 20--30 need be considered, although such
large sizes cannot be excluded. From eq.\ \ref{eq:Rd}, a conservative upper
bound on the torus outer radius is then $\Ro < 12\,L_{45}^{1/2}$ pc, where
$L_{45} = L/\E{45}\,\erg$. These compact dimensions have important implications
for the dynamics because they place the torus inside the region where the black
hole gravity dominates over the galactic bulge. If the black hole mass is
\MBH\x\E7\,\Mo\ then it dominates the gravitational motions within a radius
35\,$(\MBH/\Omega^2)^{1/3}$ pc, where $\Omega$, typically of order unity, is
the rotation velocity (in \kms\,pc$^{-1}$) induced by the galactic bulge in its
interior \citep{Elitzur_Shlosman}. Since the torus is well within the black
hole sphere of gravitational influence, its dynamic origin is determined in all
likelihood by the central engine and its accretion disk, not by the accretion
from the galaxy (see also \S\ref{sec:Final} below).

%\newpage
\section{SPECTRAL INDICATORS}
\label{sec:SPP}

As noted in the Introduction, reliable analysis of the torus full SED in
individual sources requires data that are unavailable in most cases. We can
expect flux measurements at wavelengths longer than \about\ 10\,\mic\ to be
severely contaminated by the torus surroundings. Even in NGC 1068, the best
observed AGN, the validity of almost all observations at $\lambda \ge$ 10 \mic\
is questionable because the torus contributes only a fraction of the measured
flux. The situation is unlikely to improve in the foreseeable future. The
alternative approach to individual SED fitting is analysis of large data sets
in an attempt to identify statistical trends that might constrain the likely
physical range of torus parameters. Here we discuss the spectral indicators
most commonly used in such analyses.

\ifemulate
%%%%%%%%%%%%%%%%%%%%%%%%%%%%%%%%%%%%%%%%%%%%%%%%%%%%%%
\begin{figure*}[ht]
\begin{minipage}{\textwidth}
 \Figure{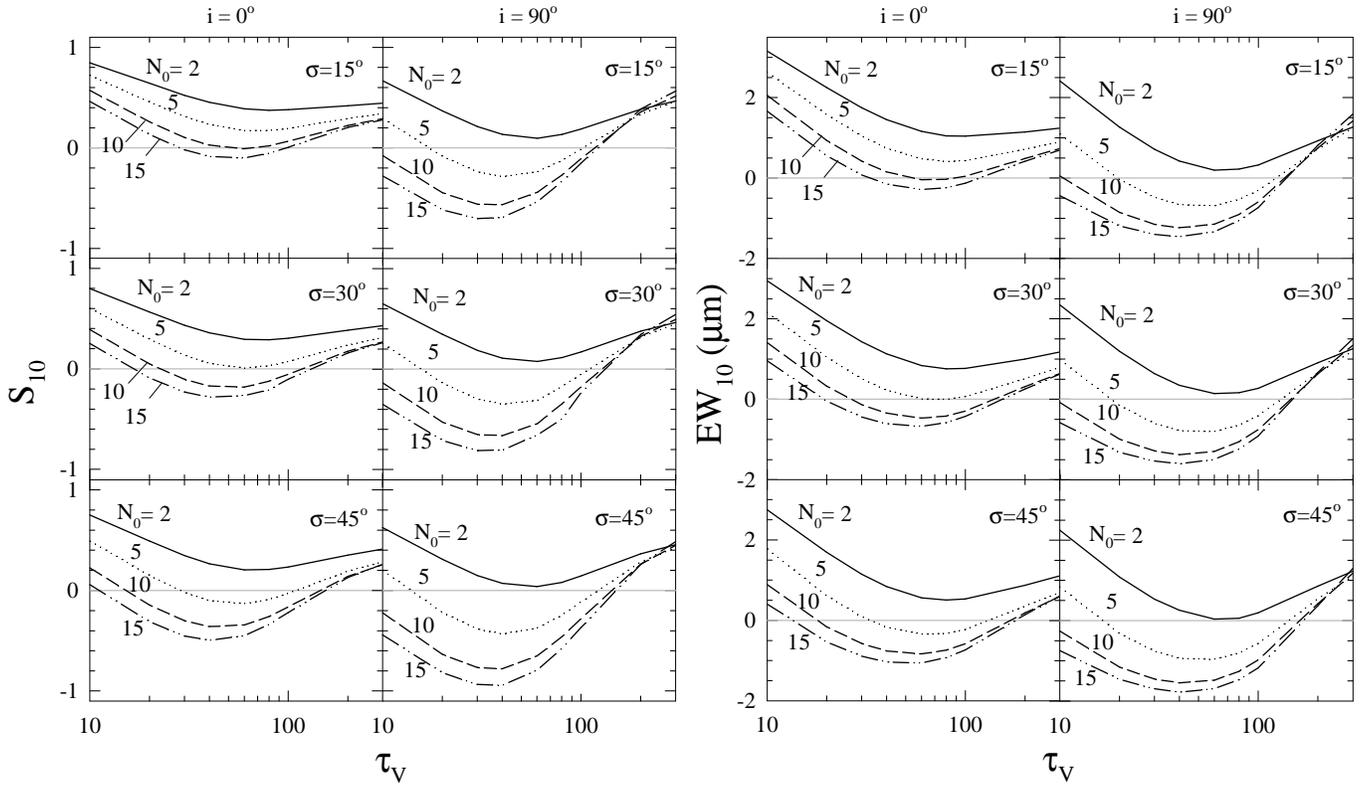}{\hsize}

\caption{Indicators of the 10 \mic\ silicate feature (see eq.\ \ref{eq:Sil10}):
variations of the feature strength ({\em left}) and equivalent width ({\em
right}) with the optical depth \tV\ of individual clouds. Model parameters are
$q$ = 2, $Y$ = 30. Other parameters as marked.  The overall optical depth along
radial equatorial rays extends all the way to \No\tV\ = 4,500 in these models,
yet the 10\mic\ absorption feature is never deep.}
 \label{Fig:Sil10a}
\end{minipage}
\end{figure*}
%%%%%%%%%%%%%%%%%%%%%%%%%%%%%%%%%%%%%%%%%%%%%%%%%%%%%%%%%%%%%%%%%

%%%%%%%%%%%%%%%%%%%%%%%%%%%%%%%%%%%%%%%%%%%%%%%%%%%%%%
\begin{figure*}[ht]
\begin{minipage}{\textwidth}
 \Figure{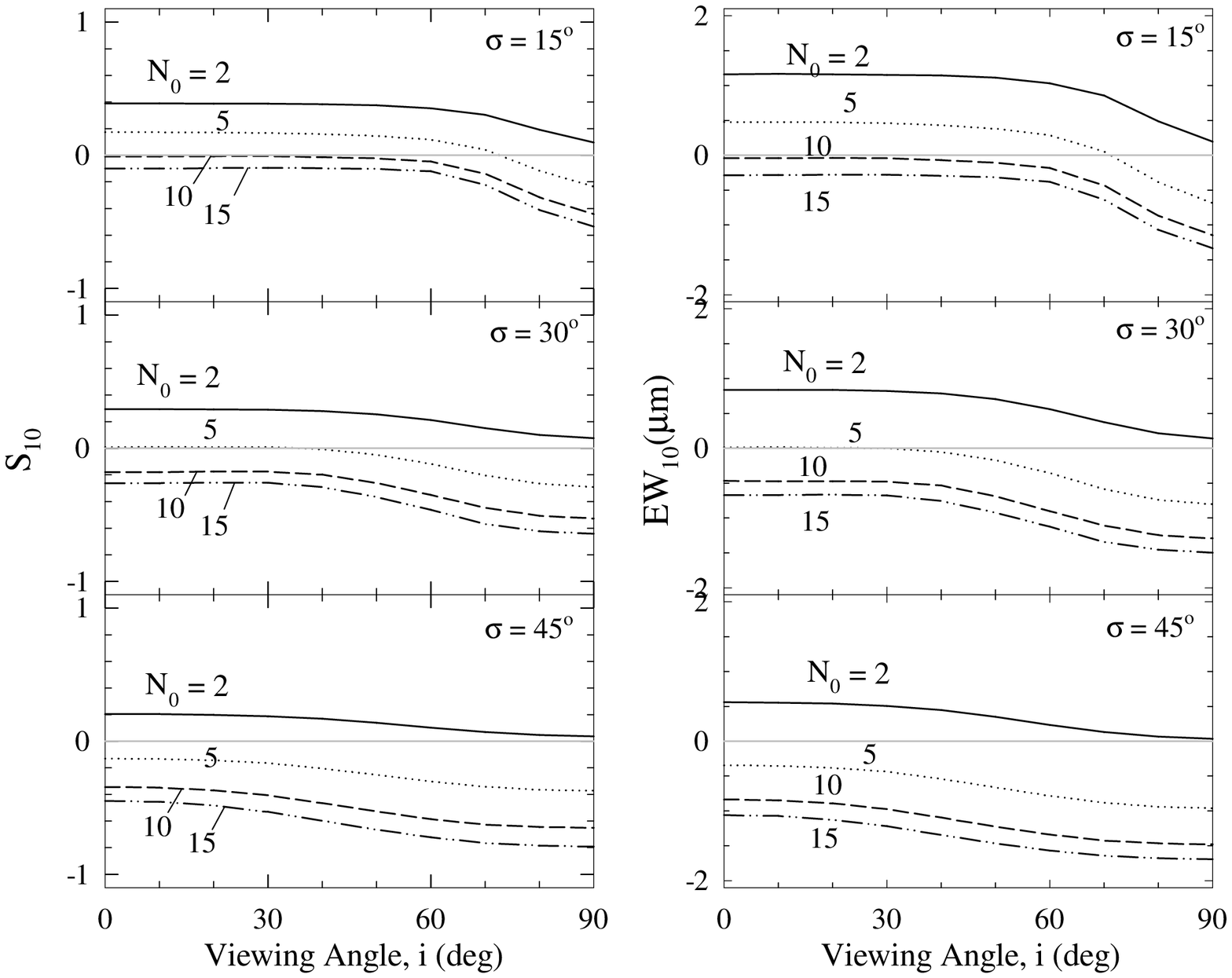}{0.8\hsize}

\caption{Variation of the 10 \mic\ feature strength ({\em left}) and equivalent
width ({\em right}) with viewing angle. Model parameters are \tV\ = 60 $q$ = 2,
$Y$ = 30. Other parameters as marked.}
 \label{Fig:Sil10b}
\end{minipage}
\end{figure*}
%%%%%%%%%%%%%%%%%%%%%%%%%%%%%%%%%%%%%%%%%%%%%%%%%%%%%%%%%%%%%%%%%
\fi

\subsection{The Silicate 10 \mic\ Feature}
\label{sec:sil}

Amorphous silicate grains have strong opacity peaks due to the Si--O stretching
and the O--Si--O bending modes, leading to broad features around 10 and 18
\mic\ in the observed dust radiation. The 10 \mic\ feature, the stronger of the
two, is a common analysis tool. In smooth-density torus models the 10 \mic\
feature appears in emission for face-on viewing and in absorption in edge-on
viewing. As shown in \S\ref{sec:SEDs}, clumpy models produce more elaborate
patterns. For detailed analysis of the features we fit a smooth curve, \Fcont,
to the underlying continuum of the entire spectral region by a spline
connecting the intervals 5--7 \mic, 14--14.5 \mic, and 25--31.5 \mic. Detailed
radiative transfer calculations verify that this interpolation procedure
properly reproduces the emission that would be generated by dust stripped of
its silicate features \citep{Sirocky08}. Figure \ref{Fig:silfeature} shows the
silicate feature profiles produced in some representative models. As noted
above, a 10 \mic\ emission feature emerges in edge-on viewing at \tV\ $\ga$
100. This peculiarity arises because individual clouds become optically thick
across the entire feature. The radiation emerging at this spectral range is
then dominated by emission from the bright faces of clouds on the torus far
side escaping through clear lines of sight. The effect becomes more pronounced
as \No\ decreases.

The 10 \mic\ feature peaks at 10.0 \mic\ in the absorption coefficients from
\cite{OHM}, and radiative transfer effects introduce occasional small shifts
(no larger than 0.5 \mic, mostly toward shorter wavelengths) around this value
in the emerging spectra. To quantify the feature's strength and width we
introduce two indicators:
\eq{\label{eq:Sil10}
    S_{10} = \ln{F_\lambda\over \Fcont}, \qquad
    EW_{10} = \int_{7\mu{\rm m}}^{14\mu{\rm m}}
              {F_\lambda - \Fcont\over\Fcont} d\lambda
}
The feature strength \Ssil\ is evaluated at the extremum near 10.0 \mic\ of the
continuum-subtracted spectrum. Positive values of \Ssil\ indicate an emission
feature, negative values an absorption feature. Delineating the feature from
noise in the data requires a certain minimum for the equivalent width
$EW_{10}$, depending on the detection system. Our sign definitions are matched
for both indicators so that absorption produces a negative $EW_{10}$, the
opposite of the standard.

Figure \ref{Fig:Sil10a} displays the variations of \Ssil\ and $EW_{10}$ with
the single cloud optical depth \tV\ for pole-on and edge-on viewing and various
model parameters that bracket the likely range in AGN tori. If the feature
width were the same in all models, \Ssil\ and $EW_{10}$ would be equivalent to
each other\footnote{If the feature is parametrized as $a\Fcont
e^{-(\delta\lambda/\Delta)^2}$, where $\delta\lambda$ is wavelength shift from
the peak, then \Ssil = ln(1 + $a$) while $EW_{10} = \sqrt{\pi}\,a\Delta$.}, but
because of variations in the feature shape (see figure \ref{Fig:silfeature}),
$EW_{10}$ contains independent information. Figure \ref{Fig:Sil10a} shows that
pole-on viewing produces an emission feature only for a limited set of
parameters. Clouds heated indirectly do not produce the emission feature when
$\tV\ \ga 20$ (sec.\ 3.2, part I). The radiation from a directly illuminated
cloud displays the feature in emission only toward directions with a view of a
sufficiently large fraction of the cloud's illuminated face (see fig.\ 12, part
I). An observer along the pole of a toroidal distribution will detect an
emission feature only from direct viewing of clouds located within $\beta <$
45\deg\ from the equator. Such clouds will be obscured by foreground clouds in
most cases, except when the torus width is small ($\sigma$ = 15\deg) or the
overall optical depth of the clumpy medium is small (small \No\ and \tV).
Therefore, at $i$ = 0\deg, only a small region of parameter space produces
models with a weak emission feature while most other parameters produce either
a featureless SED or a weak absorption feature. It is also evident from figure
\ref{Fig:Sil10a} that edge-on viewing is insensitive to the angular thickness
$\sigma$. Irrespective of optical depth, {\em the absorption feature produced
by a clumpy torus is never deep}. Figure \ref{Fig:Sil10b} shows the variation
of the two indicators with viewing angle for one value of \tV, a likely
representative of actual torus clouds.

Comparison with observations is hampered by the angular resolution problem. In
the \cite{Mason06} observations of NGC1068, the feature strength in the central
0\farcs4, presumably dominated by the AGN torus, is $\Ssil\ = -0.4$. Scanning
along the ionization cones in 0\farcs4 steps shows large variations in \Ssil\
and a strong asymmetry in its spatial distribution. Measurements with larger
apertures contain significant  contribution from the ionization cones, and {\em
Spitzer} observations may be further contaminated by still larger dusty
structures. Nevertheless, when these observations produce clear differences
between type 1 and type 2 sources, it seems reasonable to attribute such global
trends to differences in viewing angles and to compare our model results with
the observed trends while considering the actual numerical values only as
guidance. The most detailed data come from the recent compilation of {\em
Spitzer} mid-IR spectra by \cite{Hao07}. Although a loosely defined sample, it
is the largest gathered thus far, including 24 type 1 quasars, 45 Seyfert 1 and
47 Seyfert 2 galaxies. The QSOs display almost exclusively an emission feature
with $0.45 \ge \Ssil \ge 0.05$, but the Seyfert 1 galaxies are clustered around
zero feature strength, occupying the range $0.35 \ge \Ssil \ge -0.25$. Almost
all Seyfert 2 galaxies display the 10 \mic\ feature in absorption, with the
distribution showing a strong peak at $-0.1 \ge \Ssil \ge -0.4$. In addition to
the \citeauthor{Hao07}\ results, an intriguing recent development comes from
the {\em Spitzer} observations of seven high-luminosity type 2 QSOs by
\cite{Sturm06}. Although the individual spectra appear featureless, the sample
average spectrum shows the 10\,\mic\ feature in {\em emission}. More recently,
\cite{Polletta08}  did find the feature in absorption in a larger sample of
mid-IR selected obscured QSOs, while \cite{Weedman06} found the 10\,\mic\
feature either in absorption or absent in a sample of X-ray and mid-IR selected
obscured AGN.

A striking characteristic of all AGN spectra is the absence of any deep 10
\mic\ absorption features. Given the large optical depths implied by the X-ray
data, smooth dust models predict very deep absorption features. Shallow
absorption features are a hallmark of clumpy dust distributions irrespective of
geometry (part I), and the mild absorption strengths evident in our model
results reflect this general property. In contrast, ULIRGs display features
that reach extreme depths \citep{Hao07}. This different behavior can be
attributed to deep embedding in a dust distribution that is smooth, rather than
clumpy \citetext{\citealt{Levenson07}; see also \citealt{Spoon07},
\citealt{Sirocky08}}. In principle, cold foreground screens intercepting the
intrinsic IR emission of ULIRGs could also account for the deep silicate
absorption in these sources. However, such an explanation would require two
distinct dust components: a very optically thick dust blanketing the primary
radiation source and reprocessing its intrinsic radiation to emerge at the
enormous IR luminosities that identify ULIRGs, and an additional foreground
screen that absorbs the reprocessed IR radiation to produce the deep silicate
absorption. To remain cold, the foreground screen cannot provide the main
reprocessing of the huge intrinsic luminosity, yet it must always be aligned
along the line of sight with the primary dust blanket. Furthermore, the aligned
screens have to be selectively associated with ULIRGs identified with LINER-
and H\,II-like features because, unlike AGN, these sources never show shallow
absorption \citep[see fig.\ 11 in][]{Sirocky08}. Such screens present a
contrived solution for the 10\mic\ absorption in ULIRGs. In contrast, a single
entity of smooth-density embedding dust that is both geometrically and
optically thick accounts naturally for the total IR characteristics of deeply
absorbed ULIRGs.

Our calculations show that clumpy tori with \No\ = 2 never produce an
absorption feature and thus are ruled out for Seyfert galaxies, though perhaps
not for quasars (see \S\ref{sec:receding} below). The properties of the 10
\mic\ feature found in Seyfert galaxies are reproduced by our models for \No\
\about\ 5--15, $\sigma$ \about 15\deg--45\deg\ and \tV\ \about 30--100. When
\tV\ increases above \about\ 100, these models produce at equatorial viewing a
weak 10 \mic\ emission feature with a small equivalent width, offering a
potential explanation for the \citeauthor{Sturm06}\ finding in QSO2: the small
equivalent width would make it hard to discern the feature in individual
sources, bringing it out of the noise only in composite spectra. Therefore, if
this finding is verified it could indicate that the optical depths of torus
clouds perhaps are larger in QSOs than in Seyfert galaxies. However, this is
not a unique interpretation. Another possible explanation is that the cloud
number \No\ decreases as the luminosity increases. This point is discussed
further in \S\ref{sec:receding} below.

\subsubsection{Apparent Optical Depth}

The overall optical depth at visual along a radial ray in the torus equatorial
plane is \No\tV. With the standard dust properties employed here, the magnitude
of the optical depth at 10\mic\ is $\tau_{10} = 0.07\No\tV$. Another quantity
frequently employed in data analysis of absorption features is the apparent
optical depth at maximum absorption, obtained from $I = e^{-\tau_{\rm app}}$
where $I$ is the residual intensity. Therefore, from eq.\ \ref{eq:Sil10},
$\tau_{\rm app,10} = - S_{10}$. When the absorption is by a cold foreground
screen that does not emit itself at these wavelengths, $\tau_{\rm app,10}$ is
the actual 10\mic\ optical depth of the screen. But when the absorption arises
from a temperature gradient in the emitting dust, $\tau_{\rm app,10}$ can
differ substantially from the actual optical depth, and the dependence of the
two quantities on the dust column may bear little resemblance to each other.
This is especially true of the torus emission. As is evident from figures
\ref{Fig:Sil10a} and \ref{Fig:Sil10b}, the relation between $\tau_{10}$, the
actual optical depth, and $\tau_{\rm app,10}$ is multi-valued. Furthermore,
although $\tau_{10}$ exceeds 300 in these figures, $\tau_{\rm app,10}$ is never
larger than unity. The apparent optical depth $\tau_{\rm app,10}$ is a poor
indicator of the actual optical depth.

\subsection{Color Analysis}
\label{Sec:ColAn}

\ifemulate
%%%%%%%%%%%%%%%%%%%%%%%%%%%%%%%%%%%%%%%%%%%%%%%%%%%%%%%
\begin{figure}
 \Figure{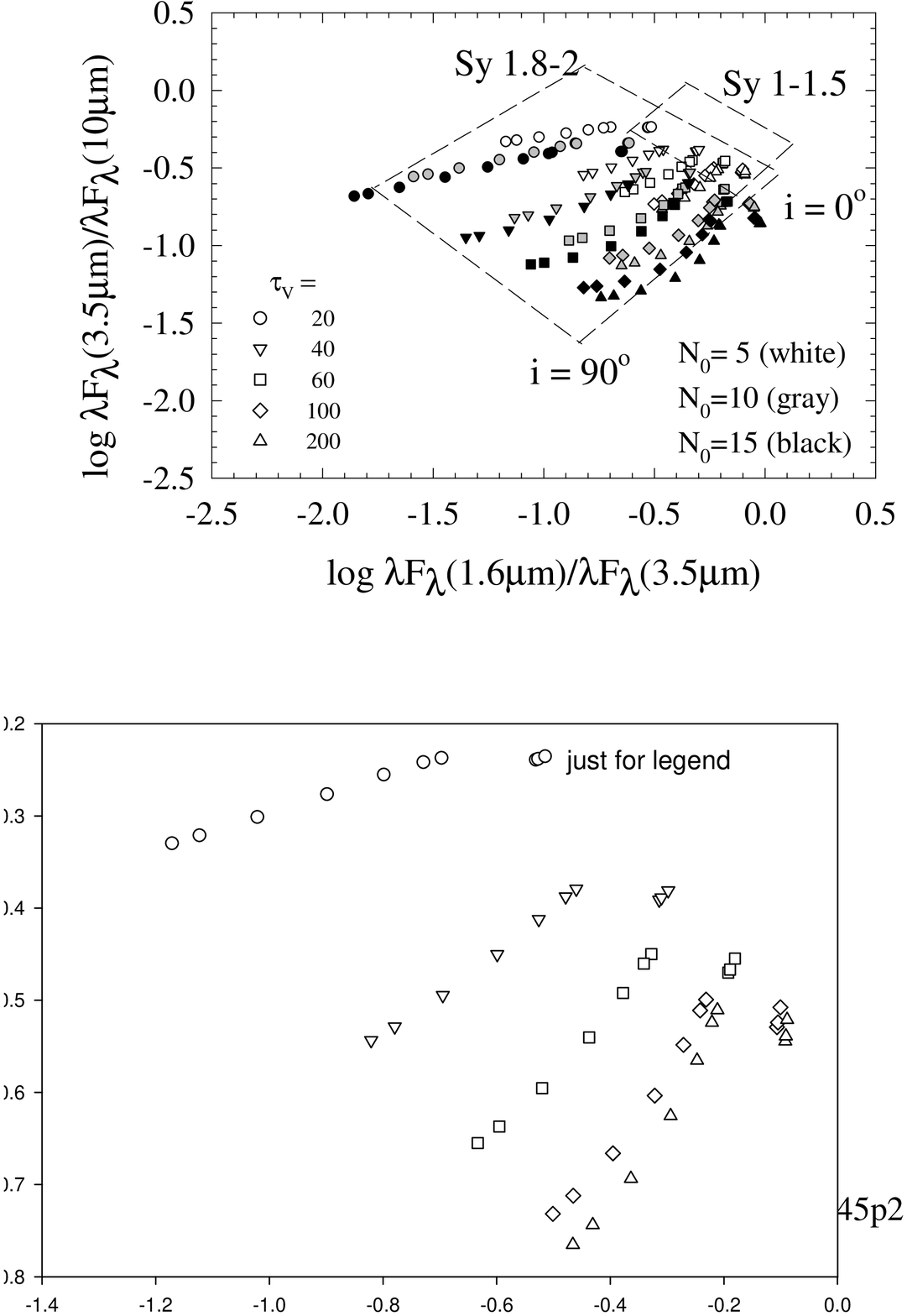}{\figsize}

\caption{Data and model results for a color-color diagram.  Dashed lines
outline the areas occupied by type 1 and type 2 sources in the
\cite{Almudena03} expanded CfA sample of Seyfert galaxies. Models have $Y =
30$, $q = 2$, $\sigma = 45\deg$, and \tV\ and \No\ as coded, respectively, with
symbols and shades. The AGN flux is added to the torus emission (type 1 model
spectrum) whenever the probability for direct view of the center exceeds
$50\%$. Each model produces a track. Positions along the track correspond to
viewing angles, varying in steps of 10\deg\ from $i$ = 0\deg\ on the right to
$i$ = 90\deg\ on the left.
}\label{Fig:colors}
\end{figure}
%%%%%%%%%%%%%%%%%%%%%%%%%%%%%%%%%%%%%%%%%%%%%%%%%%%%%%%
\fi

Color-color plots, showing correlations between two colors, are a useful way to
separate objects with similar types of spectra and reveal underlying physical
similarities. \cite{Almudena03} present data for nuclear fluxes from visual to
16 \mic\ for an expanded set of the CfA sample of Seyfert galaxies. Removing
all known sources of bias in the original CfA selection, they have constructed
what is arguably the most complete sample of AGN currently available. Torus
observations at wavelengths up to 10 \mic\ are likely to be less contaminated
by emission from the surroundings. From the \cite{Almudena03} data we find that
fluxes at 1.6 \mic, 3.5 \mic\ and 10 \mic\ provide a useful set of colors for
comparison with our model results. Compared with other combinations, the models
separate better with this choice of colors because the spectral slopes change
the most around the selected wavelengths. Figure \ref{Fig:colors} shows colors
for sets of torus models with $\sigma = 45\deg$, $Y = 30$, $q = 2$ and various
combinations of \tV\ and \No. The AGN flux is added to the torus emission
whenever the probability for direct view of the nucleus exceeds 50\%. In each
case the colors depend on the viewing angle, resulting in a track of model
results. Colors corresponding to type 1 viewing populate the upper right end of
the track, type 2 viewing the lower left. Model parameters that explain the
observations of the 10\mic\ feature also give good qualitative agreement with
the data from \cite{Almudena03}, which fall inside the two regions delineated
with dashed lines in the figure. While type 2 models are spread out along the
track, type 1 are grouped together more closely at the upper end since their
spectra are dominated by the AGN continuum and thus are similar despite the
broad range of parameters.

%\newpage
\section{ADDITIONAL IMPLICATIONS OF CLUMPINESS}
\label{sec:Others}

Comparison with IR observations shows that the likely range for optical depths
of individual torus clouds is \tV\ \about\ 30--100 and there are \No\ \about\
5--15 clouds, on average, along radial equatorial rays. Assuming standard
dust-to-gas ratio, the column density of a single cloud is \NH\ \about\
\E{22}--\E{23} cm$^{-2}$ and the torus equatorial column density is \Ntor\ =
\No\NH\ \about\ \E{23}--\E{24} cm$^{-2}$. Taking account of the torus
clumpiness has immediate implications for a number of other issues not directly
related to its IR emission.

\subsection{The Torus Mass}
\label{sec:Mass}

As shown in \S2.3 of part I, the total mass in torus clouds can be written as
$\Mtor = \mH\,\NH\int\!\!\Nc(r,\beta)\,dV$; note that \Mtor\ does not involve
the volume filling factor. With the cloud distribution from eq.\ \ref{eq:Nc}
and taking for simplicity a sharp-edge angular distribution, so that the
integration is analytic, the torus mass is $\Mtor =
4\pi\mH\,\sin\sigma\,\Ntor\Rd^2 Y\, I_q(Y)$, where $I_q$ = 1, $Y/(2\ln Y)$ and
$\frac13 Y$ for $q$ = 2, 1 and 0, respectively. Taking \Rd\ from eq.\
\ref{eq:Rd}, the mass ratio of the torus and the central black hole is
\eq{\label{eq:Mtor}
  {\Mtor\over M_\bullet} = 2\x\E{-4} {L\over\LEdd} \sin\sigma\Ntor_{,23}\,
                           Y\,I_q
}
where \LEdd\ is the Eddington luminosity and $\Ntor_{,23}$ is the equatorial
column density in \E{23} cm$^{-2}$. Since the radial thickness $Y$ is likely
$\la$ 20--30 (\S\ref{sec:size}), the torus mass is always negligible in
comparison with $M_\bullet$ when $q$ = 2. If the radial cloud distribution is
flatter, eq.\ \ref{eq:Mtor} may constrain the torus properties to keep its mass
below that of the black-hole.

\subsection{Total Number of Clouds}

As shown in part I, the total number of clouds, \ntot, is the only torus
property whose estimate involves the cloud size \Rc. Equivalently, \Rc\ can be
replaced by the volume filling factor $\phi$, since inserting eq.\ \ref{eq:Nc}
into eq.\ 3 of part I yields \Rc\ = $\phi$\Rd/\No\ at the torus inner edge. If
$\phi$ is constant throughout the torus then $\ntot \simeq \N_0^3/\phi^2$ for
the $1/r^2$ distribution, independent of the torus radial thickness $Y$.
For example, if the volume filling factor is 10\%, in order to encounter
\No\ = 5--10 clouds along each radial equatorial ray, the torus must contain
\ntot\ $\simeq$ \E4--\E5 clouds,

\ifemulate
%%%%%%%%%%%%%%%%%%%%%%%%%%%%%%%%%%%%%%%%%%%%%%%%%%%%%%%
\begin{figure*}% [ht]
\begin{minipage}{\textwidth}
 \Figure{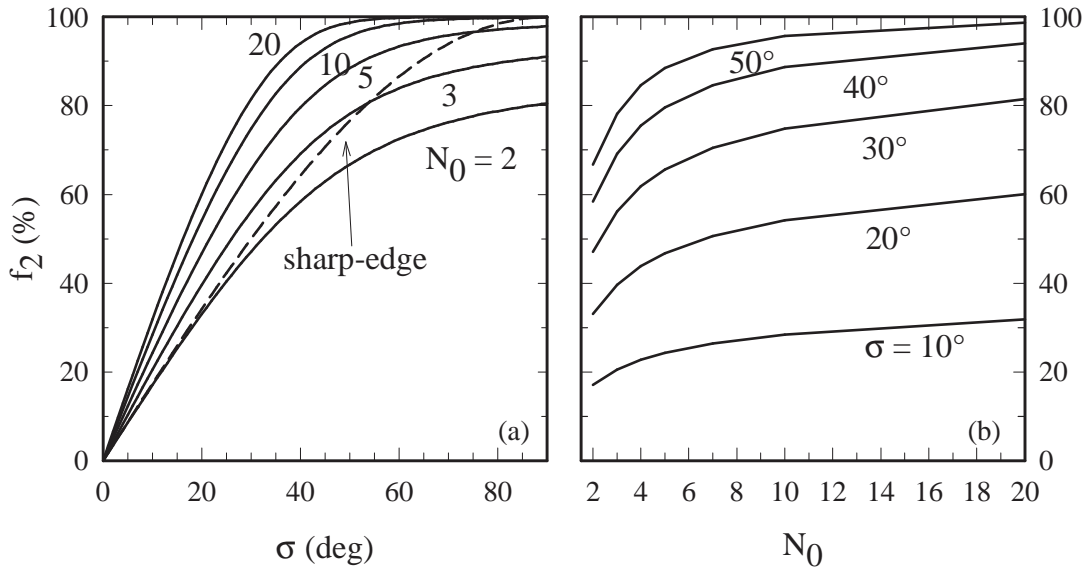}{0.8\hsize}

\caption{AGN statistics: The fraction $f_2$ of obscured sources for a clumpy
torus with Gaussian angular distribution as a function of ({\em a}) the torus
width parameter $\sigma$ and ({\em b}) the cloud number \No. The fraction
decreases when either $\sigma$ decreases at a fixed \No\ or \No\ decreases at a
fixed $\sigma$. The dashed line in panel (a) is for a clumpy torus with
sharp-edged angular profile and \No\ $\ga$ 3--4.This curve describes also the
fraction $f_2$ for every smooth-density torus, whatever its angular
distribution.
} \label{Fig:f2}

\end{minipage}
\end{figure*}
%%%%%%%%%%%%%%%%%%%%%%%%%%%%%%%%%%%%%%%%%%%%%%%%%%%%%%%%%%%%%%%%%
\fi

%  To have f2 = 70% need
%
%  No        sigma
%  2           55
%  4           40
%  5           33
%  7           30
%  10          27
%  20          25

\subsection{AGN Unification}
\label{sec:unification}

The classification of AGN into types 1 and 2 is based on the extent to which
the nuclear region is visible. In its standard formulation, the unification
approach posits the viewing angle as the sole factor in determining the AGN
type, and this is indeed the case for any smooth-density torus whose column
density declines with angle $\beta$ away from the equatorial plane. The AGN is
obscured from directions that have $e^{-\tau_{\rm V}(\beta)} \gg 1$ and visible
from those with $e^{-\tau_{\rm V}(\beta)} \ll 1$. Because of the steep
variation of $e^{-\tau}$ with $\tau$, the transition between these two regions
is sharp, occurring around the direction where $\tau_{\rm V}(\beta) = 1$.
Denote this angle $\sigma$ then, so long as $\tau_{\rm V}(0) \gg 1$ and
$\tau_{\rm V}(\half\pi) \ll 1$, all AGN viewed at ${0 \le i < \half\pi -
\sigma}$ appear as type 1 sources, those at $\half\pi - \sigma \le i \le
\frac12\pi$ as type 2. If $f_2$ denotes the fraction of type 2 sources in the
total population then $f_2 = \sin\sigma$ for all smooth-density tori,
irrespective of their specific angular profiles. This relation has been
employed in all studies of source statistics performed to date. From statistics
of Seyfert galaxies \cite{Schmitt01} find that $f_2 \simeq$ 70\%, hence their
estimate $\sigma \simeq$ 45\deg. The issue is currently unsettled because
\cite{Hao05b} have recently found that $f_2$ is only about 50\% in Seyfert
galaxies, or $\sigma \simeq$ 30\deg.

Within the clumpy torus paradigm, the difference between types 1 and 2 is not
truly an issue of orientation but of probability for direct view of the AGN.
Since that probability is always finite, type 1 sources can be detected from
what are typically considered type 2 orientations, even through the torus
equatorial plane: if \No\ = 5, for example, the probability for that is $e^{-5}
= 1/148$ on average. This might offer an explanation for the few Seyfert
galaxies reported by \cite{Almudena03} to show type 1 optical line spectra
together with 0.4--16 \mic\ SED that resemble type 2. Conversely, if a cloud
happens to obscure the AGN from an observer, that object would be classified as
type 2 irrespective of the viewing angle. In cases of such single cloud
obscuration, on occasion the cloud may move out of the line-of-sight, creating
a clear path to the nucleus and a transition to type 1 spectrum. The time scale
for such an event is determined by the cloud size and velocity. Neither
quantity can be found from the SED since optical depth is the only property of
a single cloud that can be determined from SED analysis. However, at a distance
$r_{\rm pc}$ (in pc) from a black-hole with mass \E7\MBH\ (in \Mo), the local
Keplerian speed is $208(\MBH/\rpc)^{1/2}$ km\,s$^{-1}$ and resistance to tidal
sheer implies that the size of a cloud with column density \E{23}$N_{\rm H,23}$
cm$^{-2}$ is restricted to $\la \E{16}N_{\rm H,23}r_{\rm pc}^3/\MBH$
\citep[e.g.,][]{Elitzur_Shlosman}. The ratio of this cloud size and local
Keplerian speed produces a time scale of $17\, N_{\rm H,23}r_{\rm
pc}^{3.5}/M_{\bullet7}^{1.5}$\,years, an order-of-magnitude estimate for a
cloud crossing time across the line of sight. Although the likelihood of
catching such crossing by chance is small, transitions between type 1 and 2
line spectra have been observed in a few sources \citep[see][ and references
therein]{Aret99}, and \cite{Goodrich89, Goodrich95} has argued that a couple of
these cases are consistent with the change in reddening expected from cloud
motion across the line of sight.  It is worthwhile conducting monitoring
observations in an attempt to detect additional such transitions. The most
promising candidates would be obscured systems with relatively small X-ray
obscuring columns, which may minimize the number of clouds along the line of
sight, small torus sizes, i.e., lower luminosities, and large black-hole
masses.

Accounting for the torus clumpiness, the fraction of type 2 sources is $f_2 = 1
- \int_0^{\pi/2}e^{-{\cal N_{\rm T}}(\beta)} \cos\beta d\beta$ (eq.\ 9, paper
I). The sharp-edge clumpy torus has $f_2 = (1 - e^{-{\cal N}_0})\sin\sigma$,
practically indistinguishable from a smooth-density torus when $\No$ exceeds
\about\ 3--4. However, the situation changes fundamentally for soft-edge
distributions because at every viewing angle, the probability of obscuration
increases with the number of clouds. As is evident from figure \ref{Fig:f2},
the Gaussian distribution produces a strong dependence on \No\ and significant
differences from the sharp-edge case. Since the sharp-edge angular distribution
is ruled out by observations (\S\ref{sec:geometry}), {\em the fraction of
obscured sources depends not only on the torus angular width but also on the
average number of clouds along radial rays}. While the fraction $f_2 = 70\%$
requires $\sigma = 45\deg$ in the sharp-edge case, in a Gaussian clumpy torus
it implies $\sigma = 33\deg$ when \No\ = 5 and $\sigma = 27\deg$ when \No\ =
10; in terms of the torus height and radius, $H/R$ (= tan$\sigma$) is reduced
from \about\ 1 to \about\ 0.7. It is noteworthy that the behavior of the
10\mic\ feature in the $\sigma = 30\deg$ models comes closest to matching the
observed averages of both type 1 and type 2 AGN, as is evident from figure
\ref{Fig:Sil10a}.

%\newpage
\subsection{A Receding Torus?}
\label{sec:receding}

The fraction $f_2$ of obscured AGN decreases when the bolometric luminosity
increases. This has been verified in a large number of observations that
estimate the luminosity dependence of either $f_2$ or $f_1$ (the fraction of
unobscured sources), or differences between the luminosity functions of type 1
and 2 AGN \citep[see][]{Hao05b, Simpson05, Maiolino07a}. As is evident from
figure \ref{Fig:f2}, the observed decrease of $f_2$ when $L$ increases can be
produced by either a decrease of $\sigma$ at constant \No\ or a decrease of
\No\ at constant $\sigma$. Both options are equally plausible because the torus
inner radius increases as $L^{1/2}$ (eq.\ \ref{eq:Rd}). The decreasing-$\sigma$
option would arise if the torus height is independent of luminosity or
increases more slowly than $L^{1/2}$, the decreasing-\No\ option would arise if
the torus outer radius is independent of luminosity or increases more slowly
than $L^{1/2}$.

The observed trend of $f_2$ to decrease with $L$ may arise from either $\sigma$
or \No\ or both. Source statistics cannot distinguish between the various
possibilities, the only way to decide between them is to find $L$-dependence in
other observable quantities. The 10 \mic\ silicate feature offers such an
indicator (\S\ref{sec:sil}). Among type 1 AGN,  quasars consistently produce an
emission feature but Seyfert galaxies are featureless on average, displaying
either weak emission or absorption scattered around zero feature strength. In
type 2 AGN the feature switches from clear absorption in Seyfert galaxies to
apparent emission in QSO2. That is, in both type 1 and type 2 AGN the feature
moves toward emission with the increase from Seyfert to quasar luminosities. As
is evident from figure \ref{Fig:Sil10a}, the decreasing-\No\ option naturally
produces such a universal trend: the feature appears in emission for both
pole-on and edge-on viewing when \No\ decreases to \about\,2 at a fixed
$\sigma$. In contrast, the decreasing-$\sigma$ option produces the observed
trend toward stronger emission feature only in type 1 AGN, not in type 2;
varying $\sigma$ has virtually no effect on the 10\mic\ feature in type 2
viewing. Explaining the switch toward apparent emission feature in QSO2 would
require that in this case higher luminosities not only reduce $\sigma$ but are
also accompanied by an increase in the optical depth of individual clouds.

As is evident from this discussion, current observations, if accepted at face
value and assuming that the torus contribution dominates the 10 \mic\ spectral
range on average, can be explained if an increasing luminosity causes a
decrease in the number of clouds \No. Whether or not this is also accompanied
by a decrease in the torus angular width cannot be ruled in or out. Obscuration
statistics and the 10\mic\ feature do not yet provide decisive information to
uniquely constrain the behavior of the torus parameters with increasing
luminosity.

The decreasing-$\sigma$ scenario is known as the receding torus model, first
suggested by \cite{Lawrence91}. It is intriguing that \cite{Arshakian05} and
\cite{Simpson05} derived independently an almost identical relation $\tan\sigma
\propto L^{-0.27}$. However, both studies, as well as every other analysis of
obscuration statistics thus far, were based on sharp-edge angular obscuration.
Removing this assumption affects profoundly the foundation of the receding
torus model because the dependence on the number of clouds necessitates
analysis with two free parameters, therefore $\sigma$ cannot be determined
without \No.

\subsection{X-rays and the AGN Torus}
\label{sec:X-rays}

Dusty material absorbs continuum radiation both in the UV/optical and X-rays,
therefore the dusty torus also provides X-ray obscuration. But dust-free gas
attenuates just the X-ray continuum, so clouds inside the dust sublimation
radius will provide additional obscuration only in this band.

Observations give overwhelming evidence for the orientation dependent X-ray
absorption expected from AGN unification. In general, the 2--10 keV X-ray
continuum is heavily obscured in type 2 sources and relatively unobscured in
type 1 AGN \citep[see][and references therein]{Maiolino07b}. The strong
orientation-dependence of the absorption cannot be attributed to the host
galaxy because the AGN axis, as traced by the jet position angle, is randomly
oriented with respect to the galactic disk in Seyfert galaxies \citep{Kinney00}
and the nuclear dust disk in radio galaxies \citep{Schmitt02}. Yet in spite of
the overall correspondence between the optical and X-ray obscuration, there is
a significant number of AGN for which the expected characteristics are
different in the two bands. Although substantial X-ray absorption is common
among type 2 AGN, there are also unabsorbed X-ray sources that present only
narrow emission lines in their optical spectra. Such cases can be explained
with the observational selection effect suggested by \cite{Severgnini03} and
\cite{Silverman05}: in these sources, the optical light of the host galaxy
outshines the AGN continuum and broad lines. This suggestion is supported by
the subsequent studies of \cite{Page06} and \cite{Garcet07}. The opposite case,
obscuration only in X-rays, exists too---there are type 1, broad line AGN with
significant X-ray absorption \citep{Perola04, Eckart06, Garcet07}. Extreme
cases include quasars whose optical spectrum shows little or no dust extinction
while their X-ray continuum is heavily affected by Compton thick absorption
\citep{Braito04, Gallagher06}. This cannot be attributed to observational
selection effects.

Obscuration that affects the X-rays but not the optical arises naturally from
absorption by dust-free clouds. Conclusive evidence for such absorption comes
from the short time scales for transit of X-ray absorbing clouds across the
line of sight, which establish the existence of obscuring clouds inside the
dust sublimation radius \citep{Risaliti02}. Extreme cases involve 4 hour
variability \citep{Elvis04} and variations in absorbing column of more than
\E{24} cm$^{-2}$ within two days, indicating Compton thick X-ray absorption
from a single cloud in the broad-lines region \citep{Risaliti07}. These
observations show that the torus extends inward beyond the dust sublimation
point to some inner radius \Rx\ $<$ \Rd. Clouds at $\Rx \le r \le \Rd$ partake
in X-ray absorption but do not contribute appreciably to optical obscuration or
IR emission because they are dust-free. Since every cloud that attenuates the
optical continuum contributes also to X-ray obscuration but not the other way
round, the X-ray absorbing column always exceeds the UV/optical absorbing
column, as observed \citep{Maccacaro82, Gaskell07}. Further, \cite{Maiolino01}
find that the X-ray absorbing column exceeds the reddening column in each
member of an AGN sample by a factor ranging from \about 3 up to \about 100,
implying that the bulk of the X-ray absorption comes from the clouds in the
dust-free inner portion of the torus. This could explain the \cite{Guainazzi05}
finding that at least 50\% of Seyfert 2 galaxies are Compton thick.

In steep radial distributions such as $1/r^2$, which seems to adequately
describe the torus dusty portion, most clouds are located close to the inner
radius. If this radial profile continued inward into the dust-free zone, that
region would dominate the X-ray obscuration---as observed. Similar to the
optical regime, the observed fraction of X-ray absorbed AGN varies inversely
with intrinsic luminosity \citep{Ueda03, Hasinger04, Akylas06}. This fraction
is usually derived from the statistics of sources that have at least one X-ray
obscuring cloud along the line of sight to the AGN, therefore it follows the
behavior plotted in figure \ref{Fig:f2} but with \No\ corresponding to the
total number of (dusty and dust-free) clouds. As the previous section shows,
either the radial thickness $\sigma$ or the cloud number \No\ could be
responsible for a decreasing $f_2$. \cite{Maiolino07a} find that the $f_2$
fractions follow similar trends with $L$ in the X-ray and optical regimes,
indicating that whichever intrinsic parameter is responsible for these trends
it might behave similarly in the dusty and dust-free portions of the torus.

\subsection{What is the Torus?}
\label{sec:Final}

In the ubiquitous sketch by \cite{Urry95}, the AGN central region, comprised of
the black hole, its accretion disk and the broad-line emitting clouds, is
surrounded by a large doughnut-like structure---the torus. This hydrostatic
object is a separate entity, presumably populated by molecular clouds accreted
from the galaxy. Gravity controls the orbital motions of the clouds, but the
origin of vertical motions capable of sustaining the ``doughnut'' as a
hydrostatic structure whose height is comparable to its radius was recognized
as a problem since the first theoretical study by \cite{Krolik88}.

Two different types of observations now show that the torus may be a smooth
continuation of the broad lines region (BLR), not a separate entity. IR
reverberation observations by \cite{Suganuma06} show that the dust innermost
radius scales with luminosity as $L^{1/2}$ and is uncorrelated with the black
hole mass, demonstrating that the torus inner boundary is controlled by dust
sublimation (eq.\ \ref{eq:Rd}), not by dynamical processes. Moreover, in each
AGN for which both data exist, the IR time lag is the upper bound on all time
lags measured in the broad lines, a relation verified over a range of \E6\ in
luminosity. This finding shows that the BLR extends all the way to the inner
boundary of the dusty torus, validating the \cite{Netzer_Laor} proposal that
the BLR size is bounded by dust sublimation. The other evidence is the finding
by \cite{Risaliti02} that the X-ray absorbing columns in Seyfert 2 galaxies
display time variations caused by cloud transit across the line of sight. Most
variations come from clouds that are dust free because of their proximity ($<$
0.1 pc) to the AGN, but some involve dusty clouds at a few pc. Other than the
different time scales for variability, there is no discernible difference
between the dust-free and dusty X-ray absorbing clouds, nor are there any gaps
in the distribution.

These observations suggest that the X-ray absorption, broad line emission and dust
obscuration and reprocessing are produced by a single, continuous distribution
of clouds. The different radiative signatures merely reflect the change in
cloud composition across the dust sublimation radius \Rd. The inner clouds are
dust free. Their gas is directly exposed to the AGN ionizing continuum,
therefore it is atomic and ionized, producing the broad emission lines. The
outer clouds are dusty, therefore their gas is shielded from the ionizing
radiation, and the atomic line emission is quenched. Instead, these clouds are
molecular and dusty, obscuring the optical/UV emission from the inner regions
and emitting IR. Thus the BLR occupies $r < \Rd$ while the torus is simply the
$r > \Rd$ region. Both regions absorb X-rays, but because most of the clouds
along each radial ray reside in its BLR segment, that is where the bulk of the
X-ray obscuration is produced. Since the X-ray obscuration region (XOR)
coincides mostly with the BLR, it seems appropriate to name this region instead
BLR/XOR. By the same token, since the unification torus is just the outer
portion of the cloud distribution and not an independent structure, it is
appropriate to rename it the TOR for Toroidal Obscuration Region. The close
proximity of BLR and TOR clouds should result in cases of partial obscuration,
possibly leading to observational constraints on cloud sizes.

The merger of the ionized and the dusty clouds into a single population offers
a solution to the torus vertical structure problem. Mounting evidence for cloud
outflow \citep[see, e.g.,][]{Elvis_winds} indicates that instead of a
hydrostatic ``doughnut'', the TOR is just one region in the clumpy wind coming
off the black-hole accretion disk \citep[see][and references
therein]{Elitzur_Shlosman}. The accretion disk appears to be fed by a midplane
influx of cold, clumpy material from the main body of the galaxy. Approaching
the center, conditions for developing hydromagnetically- or radiatively-driven
winds above this equatorial inflow become more favorable. The disk-wind
rotating geometry provides a natural channel for angular momentum outflow from
the disk and is found on many spatial scales, from protostars to AGN
\citep{Blandford_Payne, Emmering92, Ferreira07}. The composition along each
streamline reflects the origin of the outflow material at the disk surface. The
disk outer regions are dusty and molecular, as observed in water masers in some
edge-on cases \citep{Greenhill05}. At smaller radii the dust is destroyed and
the disk composition switches to atomic and ionized, producing a double-peak
signature in some emission line profiles \citep{Eracleous04}. The outflow from
the atomic and ionized inner region feeds the BLR and produces many atomic line
signatures, including evidence for the disk wind geometry \citep{Hall03}.
Clouds uplifted from the disk dusty and molecular outer region feed the TOR and
may have been detected in water maser observations of Circinus
\citep{Greenhill03} and NGC 3079 \citep{Kondratko05}. Indeed,
\cite{Elitzur_Shlosman} derive the cloud properties from constraints deduced
from clumpy models for the IR emission and find that they provide the right
conditions for H$_2$O maser action. In both the inner and outer outflow
regions, as the clouds rise and move away from the disk they expand and lose
their column density, limiting the vertical scope of X-ray absorption, broad
line emission and dust obscuration and emission. The result is a toroidal
geometry for both the BLR/XOR and the TOR. Because of the strong
photoionization heating of BLR clouds they may rise to relatively lower heights
than the TOR dusty clouds. Detailed comparisons of X-ray and optical
obscuration in individual sources and in large samples should help to constrain
the parameters \No, $\sigma$ and \tV\ separately for the TOR and the BLR/XOR.
Such comparisons must consider the large scatter of obscuration in individual
sources around the sample mean (see paper I, \S4.2).  In the outflow scenario,
the TOR disappears when the bolometric luminosity decreases below \about\
\E{42} \erg\ because the accretion onto the central black hole can no longer
sustain the required cloud outflow rate \citep{Elitzur_Shlosman, Elitzur07}.
With further luminosity decrease, suppression of cloud outflow spreads radially
inward and the BLR, too, disappears. The recent review by \cite{Ho08} presents
extensive observational evidence for the disappearance of the torus and the BLR
in low luminosity AGN.

The Circinus Seyfert 2 core provides the best glimpse of the AGN
dusty/molecular component. Water masers trace both a Keplerian disk and a disk
outflow \citep{Greenhill03}. Dust emission at 8--13\mic\ shows a disk embedded
in a slightly cooler and larger, geometrically thick torus \citep{Tristram07}.
The dusty disk coincides with the maser disk in both orientation and size. The
outflow masers trace only parts of the torus. The lack of full coverage can be
attributed to the selectivity of maser operation---strong emission requires
both pump action to invert the maser molecules in individual clouds and
coincidence along the line of sight in both position and velocity of two maser
clouds \citep{Kartje99}. Proper motion measurements and comparisons of the disk
and outflow masers offer a most promising means to probe the structure and
motion of TOR clouds.

\section{SUMMARY AND DISCUSSION}
 \label{sec:Discussion}

We have developed a formalism for handling radiative transfer in clumpy media
and applied it to the IR emission from the AGN dusty torus. In the calculations
we execute only the first two steps of the full iteration procedure outlined in
\S3.2, paper I,  and the moderate total number of clouds considered here
validates this procedure. When that number increases, the probability for
unhindered view of the AGN decreases, the role of indirectly heated clouds
becomes more prominent and eventually requires higher order iterations. Our
current calculations employ some additional simplifying approximations: The
grain mixture is handled in the composite-grain approximation, all dust is in
clouds without an inter-cloud medium and all clouds are identical. We have
already begun work on removing these assumptions and will report the results in
future publications.

In contrast with the smooth-density case, the clumpy problem is not well
defined because clouds can have arbitrary shapes, and any given set of
parameters can have many individual realizations. Our formalism invokes a
statistical approach for calculating an average behavior, and it is encouraging
that other approaches produce similar results. \cite{Dullemond05} conduct
``quasi-clumpy'' calculations in which the torus is modeled as a set of
axisymmetric rings, and compare the results with the smooth-density case. In
agreement with our conclusions they find that only smooth-density models can
produce very deep absorption feature while clumpy dust produces stronger
near-IR, broader SED and much more isotropic IR emission. \cite{Hoenig06}
employ 3D Monte carlo calculations that bypass some of our approximations. They
also treat different cloud realizations for the same global parameters,
allowing them to show the intrinsic scatter in SED due to the stochastic nature
of the problem. Their results are in agreement with ours, validating our
approach and the approximations we employ. Since the dust properties in their
calculations are from \cite{Draine84}, the 10 \mic\ feature reaches somewhat
larger strengths than in our calculations, which employ the \cite{OHM} ``cool"
dust \citetext{but are similar to our original results in \citealt{NIE02},
which also employed \citeauthor{Draine84} dust}. In spite of these differences,
\cite{Hoenig06} too find that the silicate absorption feature is never as deep
as expected for a uniform dust distribution, and obtain qualitatively similar
behavior of the silicate emission feature and overall SED shape.

The models presented here show that clumpy torus models are consistent with
current AGN observations if they contain \No\ \about 5--15 dusty clouds along
radial equatorial rays, each with an optical depth \tV\ \about 30--100.  The
cloud angular distribution should decline smoothly toward the axis, for
example, a Gaussian profile centered on the equatorial plane. Power-law radial
distributions $r^{-1}$ -- $r^{-2}$ produce adequate results. Dust grains with
optical properties of the standard Galactic mixture provide satisfactory
explanation to the IR observations. The behavior of the 10\mic\ silicate
feature, in particular the lack of any deep absorption features, is reproduced
naturally without the need to invoke any special dust properties. Several
suggestions that the abundance or composition of AGN dust might differ from its
Galactic counterpart can be discarded because of subsequent developments.
\cite{Risaliti99} note that, assuming standard dust abundance, the large column
densities discovered in X-ray absorption imply torus masses in excess of the
dynamical mass, posing a problem for the system stability. However, their mass
estimates were based on the uniform mass distribution and large torus sizes
derived from smooth-density models. The compact sizes and steep density
distributions of clumpy models eliminate the problem (see \S\ref{sec:Mass}).
\cite{Maiolino01} suggested that the widely different UV and X-ray extinctions
they found in individual sources could imply low dust abundance, but the
subsequent discovery of rapid variations shows that X-ray obscuration by
dust-free clouds is the more likely explanation (see \S\ref{sec:X-rays}). They
also invoked the lack of prominent 10\mic\ absorption features as an indication
that AGN dust is different from Galactic, but this is a natural consequence of
clumpy dust distributions (see \S\ref{sec:sil}). Intrinsic extinction curves
deduced from spectral analysis of type 1 sources \citetext{see
\citealt{Czerny07} for a recent review and a comprehensive discussion of
uncertainties} generally indicate a depletion of small grains, as could be
expected: the obscuration in type 1 sources is dominated by the dusty clouds
closest to the center and these clouds contain predominantly large grains,
which survive at the smallest distances from the AGN (see
\S\ref{sec:sublimation}). There is no compelling evidence for significant
differences between the properties of AGN and Galactic dust. Other dust
compositions are not ruled out, but nothing in the current data requires major
departures from the dust grains we use.

The close proximity of dust temperatures as different as $\ga$ 800 K and \about
200--300\,K found in interferometry around 12\mic\ cannot be explained by
smooth-density models even when they account for the individual temperatures of
grains with different sizes \citep{Schartmann05}. Clumpiness resolves this
puzzling observation because the dust on the dark side of an optically thick
cloud is much cooler than on the bright side. Thanks to the mixture of
different dust temperatures at the same radial distance, clumpy models
naturally explain the torus compact size. In spite of the high anisotropy of
its obscuration, the torus emission is observed to be nearly isotropic at
$\lambda \ga$ 12 \mic. Clumpy models resolve this puzzle too, since the
emission from a torus with radial thickness $Y$ = 10 varies little with viewing
angle. The variation is especially small if the radial distribution is $1/r^2$
or steeper, and such steep radial profiles maintain a nearly isotropic emission
even at larger torus sizes.

In addition to IR observations, clumpiness significantly impacts the analysis
of other data, in particular obscuration statistics. The fraction $f_2$ of
obscured sources is controlled not only by the torus angular thickness
$\sigma$, as in all analyses to date, but also by the cloud number \No. With
\No\ = 5, a 70\% fraction of type 2 AGN implies $\sigma$ \about\ 30\deg\
instead of the standard 45\deg. Observations indicate that increasing the
bolometric luminosity from the Seyfert to the quasar regime induces (1) a
decrease of $f_2$ and (2) a switch to emission feature at 10 \mic\ for both
type 1 and some type 2 AGN. Both trends can be explained with a change in a
single torus parameter---\No\ decreases from \about\ 5 in Seyfert galaxies to
\about\ 2 in QSO (see figures \ref{Fig:Sil10a} and \ref{Fig:f2}). Decreasing
$\sigma$, the scenario known as the receding torus model, explains the first
trend but has no effect on the second. The emergence of the 10 \mic\ in
emission would require in this case the additional increase of individual
clouds optical depth to \tV\ $\ga$ 100 in QSO.

The decreasing-\No\ scenario provides the simplest explanation for the trends
observed when $L$ is increasing, but that does not guarantee its validity. This
demonstrates the difficulties in deducing the model parameters from
observations that cannot yet resolve the torus basic ingredient---the
individual dusty clouds. The problem is compounded by the lack of angular
resolution that hinders clean separation of the torus component from the flux
measured at most IR wavelengths and by the degeneracies of the radiative
transfer solutions that prevent decisive, one-to-one associations between model
parameters and observable quantities. The only practical way around these
difficulties is to match trends identified in the data with similar general
properties of the models.

Our main conclusions can be summarized as follows:
\begin{itemize}
\parskip -3pt

\item The torus angular distribution has to be soft edged

\item Clumpy models can produce nearly isotropic IR emission together with
extremely anisotropic obscuration

\item Clumpy models can explain all current observations with compact torus
sizes; SED fitting is a poor constraint on the size

\item Standard interstellar dust describes adequately AGN observations; there
does not seem to be a need for any major modifications of grain properties

\item Clumpy sources never produce a very deep silicate feature; apparent
optical depth, obtained from $I = e^{-\tau_{\rm app}}$ where $I$ is the
residual intensity at maximum absorption, is a poor indicator of the actual
optical depth

\item The probability for direct line-of-sight to the AGN at large viewing
angles is small, but not zero

\item The statistics of obscured sources depend on both the torus angular
thickness and the number of clouds along radial rays

\item The torus and the BLR are the dusty (outer) and dust-free (inner) regions
in a continuous cloud distribution; a more appropriate designation for the
torus is Toroidal Obscuration Region (TOR)

\item X-ray obscuration comes from both TOR and, predominantly, BLR clouds

\end{itemize}

As long as IR observations are incapable of resolving individual torus clouds,
one must rely on the combined evidence for clumpy structure instead of on a
``smoking gun''. Individual TOR clouds seem to have been resolved in
observations of outflow water masers in Circinus and NGC 3079. Proper motion
measurements and comparison of these masers with their disk counterparts
provide the most promising method for probing the TOR structure and kinematics.
The Circinus AGN, whose dust emission has been resolved in VLTI observations,
is an especially attractive target for studying the dusty and molecular content
of TOR clouds.

\section*{ACKNOWLEDGMENTS}
Part of this work was performed while M.E.\ spent a most enjoyable sabbatical
at LAOG, Grenoble. We thank Almudena Alonso-Herrero, Nancy Levenson and
Maria Polletta for useful comments on the manuscript. Partial support by NSF
and NASA is gratefully acknowledged.

\appendix
\section{TECHNICALITIES}
\renewcommand\H     {\hbox{$H_{\lambda}$}}
\renewcommand\r     {\hbox{$\vec r$}}

The relevant coordinates in describing both the cloud distribution and the
source function are the cloud's radial distance $r$, angle $\beta$ from the
equatorial plane and the angle $\alpha$ between its radius vector and the
AGN--observer axis (see eq.\ \ref{eq:Nc}, and  part I figure 2 and eq.\ 8). The
torus emission requires an integration along a path inclined by the viewing
angle $i$ from the torus axis at some displacement from the center (eq.\ 5,
part I). To handle the geometry we introduce a cartesian coordinate system
centered on the AGN with $z$ toward the observer and $x$--$y$ in the plane of
the sky, with the torus axis in the $y$--$z$ plane at angle $i$ from the $z$
axis. The integration path is specified by its fixed values of $x$ and $y$, so
that the angular displacement is $(\theta_x,\theta_y) = (x/D, y/D)$ and the
angular impact parameter in brightness profiles is $\theta = \sqrt{x^2 +
y^2}/D$. The integration variable is $z$. At any point $\r\ = (x,y,z)$ along
the path, the cloud coordinates are found from
\eq{
    r^2 = x^2 + y^2 + z^2,
    \qquad
    \tan\beta = {y\sin i + z\cos i\over\sqrt{x^2 + (y\cos i - z\sin i)^2}},
    \qquad
    \cos\alpha = {z\over r}.
}

The path integration in eq.\ 5, part I, produces the intensity generated by the
cloud distribution. Since our source function calculations involve only the
first two steps of the full iteration procedure described in \S3.2, part I, we
must introduce a correction to take proper account of flux conservation. With
\pAGN\ the fraction of the AGN luminosity that gets through the torus (eq.\ 8,
part I), we calculate $\IC(x,y;i)$, the brightness map of clumpy torus emission
in the direction $i$, from
\eq{\label{eq:IC-final}
  \IC = {L(1 - \pAGN)\over 4\pi\int d\cos i \int d\lambda \int \H dxdy}\, \H,
  \qquad\hbox{where}\quad
  \H(x,y;i) = \int \Pesc_{,\lambda}(\r)\, \Sl(\r)\Nc(\r)\,dz
}
Here \Sl(\r) and \Nc(\r) are, respectively, the source function and column
density of clouds at position \r\ along the integration path, and
$\Pesc_{,\lambda}(\r)$ is the probability for a photon of frequency $\lambda$
to escape from that point through the rest of the path. The torus flux at
distance $D$ and viewing angle $i$ is calculated from $\FC(i) =
(1/D^2)\int\IC(x,y;i) dxdy$. With these expressions, the spectral shape is
determined from the first two iteration steps while ensuring that the torus
emission properly obeys flux conservation (eq.\ 17, part I).

The quantity \H\ is intrinsically a function of scaled variables, $\H =
\H(x/\Rd, y/\Rd; i)$, because the brightness at position $(x,y)$ depends only
on the distribution of dust in temperature and optical depth along the path
\citep{IE97}. Therefore, from eq.\ \ref{eq:IC-final} the brightness has the
form $\IC(\theta_x,\theta_y;i) = (L/4\pi\Rd^2)f(\theta_x/\td,\theta_y/\td)$,
where $f$ is a dimensionless function of the scaled angular displacements.
Since the brightness scale $L/4\pi\Rd^2$ is determined by the dust sublimation
temperature $T_{\rm sub}$ (eq.\ \ref{eq:Rd}) , the only effect of the
luminosity is to set the overall angular scale \td, effecting a self-similar
stretch of the brightness map. Similarly, the flux, \FC, is a product of the
bolometric flux \FAGN\ and a luminosity-independent spectral shape.

%\bibliographystyle{apj}
%\bibliography{AGN,Dust}

% To create reference list, run the two above, comment the one below.
% Once list is established, rename the existing .bbl file to .refs,
% comment the two above and run the one below.

%\input{AGN2rev1.refs}

\ifemulate{\end{document}}\fi

\clearpage

%%%%%%%%%%%%%%%%%%%% Torus - 2 types %%%%%%%%%%%%%%%%%%%%%%%%%%%%%%%%
\begin{figure}
 \Figure{f1a}{0.49\figsize} \hfill
 \Figure{f1b}{0.49\figsize}

\caption{Model geometry: Dusty clouds, each with an optical depth \tV\ at
visual, occupy a toroidal volume from inner radius \Rd, determined by dust
sublimation (eq.\ \ref{eq:Rd}), to outer radius $\Ro = Y\Rd$. The radial
distribution is a power law $r^{-q}$, the total number of clouds along a radial
equatorial ray is \No. Various angular distributions, characterized by a width
parameter $\sigma$, were considered. The angular distribution has a sharp-edge
on the left and a smooth boundary (e.g., a Gaussian) on the right.}
\label{Fig:torus}
\end{figure}
%%%%%%%%%%%%%%%%%%%%%%%%%%%%%%%%%%%%%%%%%%%%%%%%%%%%%%%%%%%%%%

\clearpage

%%%%%%%%%%%%%%% Pesc  %%%%%%%%%%%%%%
\begin{figure}
 \Figure{f2}{ \figsize}

\caption{Behavior of the probability for photon escape along a path containing
\N\ clouds (eq.\ 4 part I). {\em Upper panel}: Wavelength variation of \Pesc\
for the indicated \N\ when the single cloud optical depth is \tV\ at visual.
{\em Lower panel}: The probability for an AGN photon to escape through the
torus in direction $i$ from the pole when each cloud is optically thick; this
is also the probability for unobscured view of the AGN at viewing angle $i$.
The total number of clouds varies according to $\NT(\beta) = \
\No\exp(-\beta^2/\sigma^2)$, where $\beta = \frac12\pi - i$ is angle from the
equatorial plane, with  $\sigma = 45\deg$ and \No\ as marked.}
 \label{Fig:Pesc}
\end{figure}

%%%%%%%%%%%%%%%%%%%%%%%%%%%%%%%%%%%%%%%%%%%%%%%%%%%%%%%

\clearpage

%%%%%%%%%%%%%%% Models of sharp edge torus vs. Gaussian dsitribution %%%%%%%%%%
\begin{figure}
 \Figure{f3}{\figsize}
\caption{Model spectra for a torus of clouds, each with optical depth \tV\ =
60. Radial distribution with $q = 1$ out to $Y$ = 30, with \No\ = 5 clouds
along radial equatorial rays (see eq.\ \ref{eq:Nc}). The angular distribution is
sharp-edged in the top panel, Gaussian in the bottom one (cf.\ fig.
\ref{Fig:torus}); both have a width parameter $\sigma$ = 45\deg. Different
curves show viewing angles that vary in 10\deg\ steps from pole-on ($i$ =
0\deg) to edge-on ($i$ = 90\deg). Fluxes scaled with $\FAGN = L/4\pi D^2$.}
\label{Fig:srpGauss}
\end{figure}
%%%%%%%%%%%%%%%%%%%%%%%%%%%%%%%%%%%%%%%%%%%%%%%%%%%%%%%

\clearpage

%%%%%%%%%%%%%%% Models vs. Data for Type 1,2 %%%%%%%%%%
\begin{figure}
 \Figure{f4}{\figsize}

\caption{Observations of type 1 and type 2 sources compared with clumpy torus
model spectra. The type 1 composite data are from \cite{Sanders89},
\cite{Elvis94}, \cite{Hao07} and \cite{Netzer07}. The type 2 data are from the
following sources: a) \cite{Mason06}; b) various observations with aperture
$\le$ 0.5\arcsec\ listed in \cite{Mason06}; c) \cite{Almudena03}; d)
\cite{Prieto+04}. In the model calculations, plotted with broken lines, each
cloud has optical depth \tV\ = 30. Other parameters are $\sigma$ = 30\deg, $q$
= 0--3, as marked, $Y$ = 30 and \No\ = 5. The angular distribution in this and
all subsequent figures is Gaussian. The models in the upper panel are for
pole-on viewing ($i$ = 0\deg), in the bottom panel for edge-on ($i$ = 90\deg).
} \label{Fig:DataComp}
\end{figure}
%%%%%%%%%%%%%%%%%%%%%%%%%%%%%%%%%%%%%%%%%%%%%%%%%%%%%%%

\clearpage

%%%%%%%%%%%%%%% tauV dependence %%%%%%%%%%%%%%
\begin{figure}
 \Figure{f5}{\figsize}
\caption{Dependence of the torus SED on the single cloud optical depth \tV.
Other parameters are $\sigma = 45\deg$, \No\ = 5 and $Y$ = 30. Radial power law
with $q$ = 1 in the left panels, $q$ = 2 in the right ones. Pole-on viewing in
the top panels, edge-on at the bottom.
} \label{Fig:taudep}
\end{figure}
%%%%%%%%%%%%%%%%%%%%%%%%%%%%%%%%%%%%%%%%%%%%%%%%%%%%%%%

\clearpage

%%%%%%%%%%%%%%% Torus emission - NT dependence %%%%%%%%%%%%%%
\begin{figure}
 \Figure{f6}{\figsize}
\caption{Dependence of the torus SED on the number of clouds along a radial
equatorial ray. Each cloud has \tV\ = 60. Angular width $\sigma$ = 45\deg,
power law radial distribution with $q$ = 1 (left panels) and 2 (right),
extending to $Y$ = 30. Different curves in each panel show viewing angles that
vary from 0\deg\ (top curve) to 90\deg\ (bottom) in 10\deg\ steps.
}
 \label{Fig:SED-torus}
\end{figure}
%%%%%%%%%%%%%%%%%%%%%%%%%%%%%%%%%%%%%%%%%%%%%%%%%%%%%%%

\clearpage

%%%%%%%%%%%%%%% NT dependence with added AGN when prob.>50% %%%%%%%%%%%%%%
\begin{figure}
 \Figure{f7}{\figsize}

\caption{Model spectra as in figure \ref{Fig:SED-torus} for $q$ = 2, only with
the AGN contribution added. For each set of parameters, the probability that
the AGN emission will actually be observed can be read from the plots of \Pesc\
in the lower panel of figure \ref{Fig:Pesc}. The break in the SED at $\lambda$
= 1 \mic\ is an artifact of our parametrization of the input spectrum (see
\S3.1.1, part I).} \label{Fig:SED}
\end{figure}
%%%%%%%%%%%%%%%%%%%%%%%%%%%%%%%%%%%%%%%%%%%%%%%%%%%%%%%

\clearpage

%%%%%%%%%%%%%%%%%%%%%%%%%%%%%%%%%%%%%%%%%%%%%%%%%%%%%%%
\begin{figure}
 \Figure{f8}{\figsize}
\caption{Dependence of the torus model spectra on the width
parameter $\sigma$ of the Gaussian angular distribution. Each cloud
has \tV\ = 60. The cloud radial distribution is a power law with
\No\ = 5 and $q$ = 2 extending to $Y$ = 30. Viewing angles vary from
0\deg\ to 90\deg\ in 10\deg\ steps. } \label{Fig:SED-opang}
\end{figure}
%%%%%%%%%%%%%%%%%%%%%%%%%%%%%%%%%%%%%%%%%%%%%%%%%%%%%%%

\clearpage

%%%%%%%%%%%%%%%%%%%%%%%%%%%%%%%%%%%%%%%%%%%%%%%%%%%%%%
\begin{figure}
 \Figure{f9}{\figsize}

\caption{Dependence of the torus model spectra on the power $q$ of the radial
density distribution, which extends to $Y$ = 10 in the top panels and $Y$ = 30
in the bottom ones. \No\ = 5 clouds with \tV\ = 60 each. Angular width $\sigma$
= 45\deg. Viewing angles from 0\deg\ to 90\deg\ in 10\deg\ steps. The emission
anisotropy decreases when $q$ increases.} \label{Fig:SED-qdep}
\end{figure}
%%%%%%%%%%%%%%%%%%%%%%%%%%%%%%%%%%%%%%%%%%%%%%%%%%%%%%%

\clearpage

%%%%%%%%%%%%%%%%%%%%%%%%%%%%%%%%%%%%%%%%%%%%%%%%%%%%%%%
\begin{figure}
 \Figure{f10}{\figsize}

\caption{Variation of the torus flux with viewing angle at different
wavelengths, as marked. \No\ = 5 clouds with \tV\ = 60 each in a radial
distribution with $q$ = 1 (top panel) and 2 (bottom) extending to $Y$ = 30.
Angular width $\sigma$ = 45\deg. Since the emission is normalized to the AGN
flux, the plotted quantity provides the bolometric correction for each
displayed wavelength.} \label{Fig:Aniso-lam}
\end{figure}
%%%%%%%%%%%%%%%%%%%%%%%%%%%%%%%%%%%%%%%%%%%%%%%%%%%%%%%

\clearpage

%%%%%%%%%%%%%%%%%%%%%%%%%%%%%%%%%%%%%%%%%%%%%%%%%%%%%%%
\begin{figure}
 \Figure{f11a}{0.75\figsize}
 \Figure{f11b}{0.75\figsize}
\caption{Anisotropy indicators for the torus bolometric flux \Ftor. All models
have $q$ = 2 and $Y$ = 30. {\em Top:} Variation of \Ftor\ with viewing angle
when \tV\ = 60 for various values of \No\ (cloud number) and $\sigma$
(torus angular width), as indicated; note the changing vertical scale. {\em
Bottom:} Ratio of the torus bolometric fluxes along the axis and equator as a
function of single cloud optical depth. The torus angular width is $\sigma$ =
45\degr.}
 \label{Fig:fbol}
\end{figure}
%%%%%%%%%%%%%%%%%%%%%%%%%%%%%%%%%%%%%%%%%%%%%%%%%%%%%%%

\clearpage

%%%%%%%%%%%%%%%%%%%%%%%%%%%%%%%%%%%%%%%%%%%%%%%%%%%%%%
\begin{figure}
 \Figure{f12}{\figsize}

\caption{Dependence of the SED of a clumpy torus on the radial thickness $Y =
\Ro/\Rd$, as marked. Radial distribution with $q$ = 1 (left panels) and 2
(right). All models have \No\ = 5 clouds with \tV\ = 60 each, and $\sigma$ =
45\deg. Pole-on viewing in the top panels, edge-on at the bottom. Note that the
curves in the left-bottom panel have a similar shape at $\lambda \la 15$ \mic\
and would nearly overlap if normalized to a common wavelength in that range
instead of \FAGN. }
 \label{Fig:SED-Ydep}
\end{figure}
%%%%%%%%%%%%%%%%%%%%%%%%%%%%%%%%%%%%%%%%%%%%%%%%%%%%%%

\clearpage

%%%%%%%%%%%%%%%%%%%%%%%%%%%%%%%%%%%%%%%%%%%%%%%%%%%%%%
\begin{figure*}[ht]
\begin{minipage}{\textwidth}
 \Figure{f13}{0.9\hsize}
\caption{Fraction of the torus flux enclosed within a circle with angular
radius $\theta$ centered on the AGN. Wavelengths, in \mic, as labeled (some
labels omitted for clarity). \td\ is the angle equivalent of the torus inner
radius \Rd\ (eq.\ \ref{eq:Rd}) at the observer's location; for reference, \td =
0\farcs02 for \Rd\ = 1\,pc at 10 Mpc. All models have \No\ = 5, \tV\ = 60,
$\sigma = 45\deg$ and $q$ = 1 or $q$ = 2, as marked. In each case, pole-on
viewing in the left panels, edge-on at right. Torus sizes, from top to bottom,
are $Y$ = 10, 30 and 100. }
 \label{Fig:FracFlx}
\end{minipage}
\end{figure*}
%%%%%%%%%%%%%%%%%%%%%%%%%%%%%%%%%%%%%%%%%%%%%%%%%%%%%%%%%%%%%%%%%

\clearpage

%%%%%%%%%%%%%%%%%%%%%%%%%%%%%%%%%%%%%%%%%%%%%%%%%%%%%%%
\begin{figure}
 \Figure{f14a}{0.7\figsize} \\
 \Figure{f14b}{0.7\figsize}

\caption{Surface brightness for torus models with \tV\ = 60, \No\ = 5, $\sigma$
= 45\deg, $Y$ = 30 and $q$ = 1 and 2, as indicated. The top panel shows the
radial variation of intensity with angular displacement from the center for
pole-on viewing and a set of wavelengths as marked. The AGN emission, which is
not shown, corresponds to a narrow spike at $\theta/\td \ll 1$ (see text). Each
intensity profile is normalized to its brightness level at $\theta = \td$,
shown in the bottom panel together with the corresponding brightness
temperature.
} \label{Fig:IntProfl}
\end{figure}

%%%%%%%%%%%%%%%%%%%%%%%%%%%%%%%%%%%%%%%%%%%%%%%%%%%%%%%

\clearpage

%%%%%%%%%%%%%%%%%%%%%%%%%%%%%%%%%%%%%%%%%%%%%%%%%%%%%%%%%%%%%%%%%%%%%%%%%%%%
\begin{figure*}
\begin{minipage}{\textwidth}
 \centering \leavevmode
 \includegraphics[width=0.8\hsize,clip]{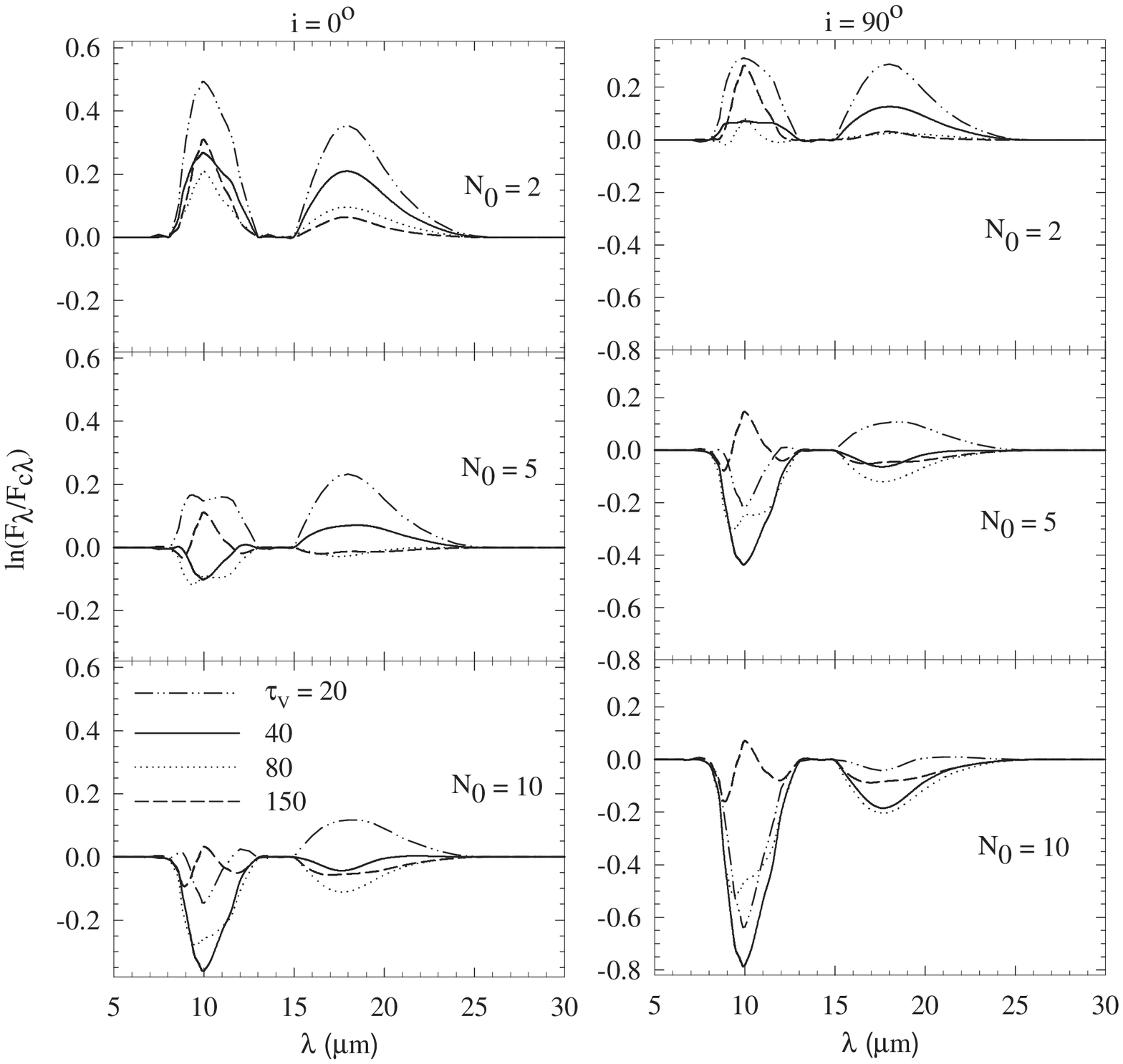}

\caption{Spectral shape of the silicate 10 and 18 \mic\ features: $F_\lambda$
is the torus emission in the 5--30 \mic\ region and \Fcont\ is the smooth
underlying continuum obtained by a spline connecting the feature-free segments
in this spectral region (see text). All models have $q$ = 2, $Y$ = 30, $\sigma$
= 45\deg, and \No\ as marked in each panel. Curves correspond to different \tV,
as labeled. Panels on the left correspond to pole-on viewing, on the right to
edge-on; note the different scales of the vertical axes in the two cases.
} \label{Fig:silfeature}
\end{minipage}
\end{figure*}
%%%%%%%%%%%%%%%%%%%%%%%%%%%%%%%%%%%%%%%%%%%%%%%%%%%%%%%%%%%%%%%%%%%%%%%%%%%%

\clearpage

%%%%%%%%%%%%%%%%%%%%%%%%%%%%%%%%%%%%%%%%%%%%%%%%%%%%%%
\begin{figure*}[ht]
\begin{minipage}{\textwidth}
 \Figure{f16}{\hsize}

\caption{Indicators of the 10 \mic\ silicate feature (see eq.\ \ref{eq:Sil10}):
variations of the feature strength ({\em left}) and equivalent width ({\em
right}) with the optical depth \tV\ of individual clouds. Model parameters are
$q$ = 2, $Y$ = 30. Other parameters as marked.  The overall optical depth along
radial equatorial rays extends all the way to \No\tV\ = 4,500 in these models,
yet the 10\mic\ absorption feature is never deep.}
 \label{Fig:Sil10a}
\end{minipage}
\end{figure*}
%%%%%%%%%%%%%%%%%%%%%%%%%%%%%%%%%%%%%%%%%%%%%%%%%%%%%%%%%%%%%%%%%

\clearpage

%%%%%%%%%%%%%%%%%%%%%%%%%%%%%%%%%%%%%%%%%%%%%%%%%%%%%%
\begin{figure*}[ht]
\begin{minipage}{\textwidth}
 \Figure{f17}{0.8\hsize}

\caption{Variation of the 10 \mic\ feature strength ({\em left}) and equivalent
width ({\em right}) with viewing angle. Model parameters are \tV\ = 60 $q$ = 2,
$Y$ = 30. Other parameters as marked.}
 \label{Fig:Sil10b}
\end{minipage}
\end{figure*}
%%%%%%%%%%%%%%%%%%%%%%%%%%%%%%%%%%%%%%%%%%%%%%%%%%%%%%%%%%%%%%%%%

\clearpage

%%%%%%%%%%%%%%%%%%%%%%%%%%%%%%%%%%%%%%%%%%%%%%%%%%%%%%%
\begin{figure}
 \Figure{f18}{\figsize}

\caption{Data and model results for a color-color diagram.  Dashed lines
outline the areas occupied by type 1 and type 2 sources in the
\cite{Almudena03} expanded CfA sample of Seyfert galaxies. Models have $Y =
30$, $q = 2$, $\sigma = 45\deg$, and \tV\ and \No\ as coded, respectively, with
symbols and shades. The AGN flux is added to the torus emission (type 1 model
spectrum) whenever the probability for direct view of the center exceeds
$50\%$. Each model produces a track. Positions along the track correspond to
viewing angles, varying in steps of 10\deg\ from $i$ = 0\deg\ on the right to
$i$ = 90\deg\ on the left.
}\label{Fig:colors}
\end{figure}
%%%%%%%%%%%%%%%%%%%%%%%%%%%%%%%%%%%%%%%%%%%%%%%%%%%%%%%

\clearpage

%%%%%%%%%%%%%%%%%%%%%%%%%%%%%%%%%%%%%%%%%%%%%%%%%%%%%%%
\begin{figure*}% [ht]
\begin{minipage}{\textwidth}
 \Figure{f19}{0.8\hsize}

\caption{AGN statistics: The fraction $f_2$ of obscured sources for a clumpy
torus with Gaussian angular distribution as a function of ({\em a}) the torus
width parameter $\sigma$ and ({\em b}) the cloud number \No. The fraction
decreases when either $\sigma$ decreases at a fixed \No\ or \No\ decreases at a
fixed $\sigma$. The dashed line in panel (a) is for a clumpy torus with
sharp-edged angular profile and \No\ $\ga$ 3--4.This curve describes also the
fraction $f_2$ for every smooth-density torus, whatever its angular
distribution.
} \label{Fig:f2}

\end{minipage}
\end{figure*}
%%%%%%%%%%%%%%%%%%%%%%%%%%%%%%%%%%%%%%%%%%%%%%%%%%%%%%%%%%%%%%%%%

\end{document}